\documentclass[aps,amssymb,amsmath, singlecolumn, pra, superscriptaddress]{revtex4}
\usepackage{graphicx}
\usepackage{epsfig}
\usepackage{dcolumn}
\usepackage{amsmath}
\usepackage{bm}
\usepackage{bbm}
\usepackage[colorlinks, citecolor=red]{hyperref}
\usepackage{color}
\usepackage{centernot}
\usepackage{float}
\usepackage[up]{subfigure}
\newcommand{\be}{\begin{equation}}
\newcommand{\ee}{\end{equation}}
\newcommand{\bc}{\begin{center}}
\newcommand{\ec}{\end{center}}
\newcommand{\bea}{\begin{eqnarray}}
\newcommand{\eea}{\end{eqnarray}}
\newcommand{\ba}{\begin{array}}
\newcommand{\ea}{\end{array}}

\begin{document}

\title{Noise-enhanced quantum transport on a closed loop using quantum walks}

\author{C. M. Chandrashekar}
\email{c.madaiah@oist.jp}
\author{Th. Busch}
\affiliation{Physics Department, University College Cork, Cork, Ireland}
\affiliation{Quantum Systems Unit, Okinawa Institute of Science and Technology Graduate University, Okinawa, Japan}


\begin{abstract}
We study the effect of noise on the transport of a quantum state from 
  a closed loop of $n-$sites with one of the sites as a sink. Using a discrete-time quantum walk dynamics, we
  demonstrate that the transport efficiency can be enhanced with noise
  when the number of sites in the loop is small and reduced when
  the number of sites in the loop grows.  By using the concept of
  measurement induced disturbance we identify the regimes in which
  genuine quantum effects are responsible for the enhanced transport.
\end{abstract}

\maketitle

\section{Introduction}
\label{intro}
Studying energy transfer or transport efficiencies in the presence of
quantum coherence has recently become an active area due to its
applications in many different scientific fields. A number of
microscopic models have been developed and their studies highlight the
possibility of quantum coherence playing an important role in, for
example, the energy and charge transfer process in mesoscopic bio-chemical
systems\,\cite{GB04,OLO08,NET10,NBT11,SMW11,RMK09,PH08} or artificial photosynthetic complexes\,\cite{GSN11a,GSN11b}. At the
same time first experimental studies have shown that quantum
coherence can play a significant role in photosynthetic
light-harvesting complexes\,\cite{ECR07,PHF10,CWW10} and in
chromophoric energy transport\,\cite{RMA09}.  Since one of the main
results found in many of these studies is the existence of enhanced transport
efficiency in the presence of noise\,\cite{RMK09,PH08,ECR07,CCD09,MRL08}, it has become clear that understanding the dynamics of noisy
quantum state transport can play a significant role in modeling and
understanding natural physical and bio-chemical processes.
Identifying models and dynamics in which the effect of the noise
contributes constructively to the transport efficiency is therefore an
important task. While the dynamics of quantum systems can be described
by different forms of evolution, here we concentrate on quantum walks,
which have been shown to be an effective model to understand, for
example, photosynthetic energy transfer\,\cite{ECR07,MRL08}.

Quantum walks evolve a particle in a series of superpositions in
position space\,\cite{Ria58,Fey86,Par88,LP88,ADZ93,FG98} and are the
quantum analogue to classical random walks. They are studied in two forms, continuous-time quantum walk and discrete-time quantum walk. The continuous-time quantum walk is directly defined on the position Hilbert space using an Hamiltonian with a transition probability of the amplitude to its neighboring position\,\cite{FG98}.  The discrete-time quantum walk is defined on a position Hilbert space along with an additional internal degree of freedom of the coin (particle) Hilbert space\,\cite{ADZ93}. 
Over the past decade, both the versions of quantum walks have emerged as an efficient tool to develop quantum algorithms\,\cite{Kem03,Amb03}, to have coherent control of atoms and Bose-Einstein condensates in optical lattices\,\cite{CL08,Cha11a}, to create topological phases\,\cite{KRB10}, to construct protocols for
quantum state transfer\,\cite{CDE04,KW11}, and to generate entanglement\,\cite{GC10} among many other applications. Many organic and bio-chemical systems can be modeled by considering a closed loop of $n$ sites, which has connections to other systems at specific positions. The dynamics of energy, charge and excitations on such a system can be  can be modeled using quantum walks. Therefore, quantum walks are well suited to serve as a framework to simulate, control and understand the 
dynamics of physical and bio-chemical systems\,\cite{DGV12}.

Transport process using quantum walk has also been studied using both, continuous-time model (see Ref.\,\cite{MB11} for a recent review on the topic) and discrete-time model\,\cite{BCG04,GAS11}.  Using discrete-time model the dynamics of the quantum walk on one-dimension with absorbing boundary conditions\,\cite{BCG04}  has been reported. For a system with internal degree of freedom, compared to the continuous-time model, discrete-time quantum walk can serve as a well-suited model to study the transport process.  In Ref.\,\cite{GAS11}, comparison of survival probability (inverse of transport probability) on a closed loop using classical transport and discrete-time quantum walk with an absorbing traps has been studied. As the motive for the authors of Ref.\,\cite{GAS11} was to compare the survival probability of the classical transport and discrete-time quantum walk, the effect of noise on the transport of quantum state through an absorbing traps was not considered in the study. Quantum systems are not free from the environment noise and hence, looking into the effect of noise on the quantum transport process is very important forward. In this article we present the transport process using a simple form of the discrete-time quantum walk evolution of a closed loop on $n-$ sites\,\cite{CSB07,BSC08,LP10} and study the effect of noise on the transport efficiency with one of the sites as a sink.

To measure the transport efficiency resulting from a discrete-time quantum walk on such a closed loop we designate one of the sites as a sink at which a certain part of the amplitude of the particle gets absorbed and might, for example, be transported to a neighbouring loop or network. Since this is a periodic
system it is clear that for infinitely long times 100\% of the wavefunction will be absorbed at the sink (we do not consider backflow out of the sink), however the efficiency at
short times will depend strongly on the sink's position and strength as well as the noise level in the system. By modeling the strength of the
environmental effects using a depolarizing and a dephasing channel, we find that for small loops any level of noise will lead to enhanced state transport. For small loop with a large number of sites and sink, however, a decreased efficiency is found. For a loop with a large number of sites, however, a decreased efficiency is found. Studying the transport efficiency as a function of time we also
identify regimes in which temporal enhancement exists. Finally, we
determine the quantumness of the transport by using measurement induced
disturbance\,\cite{L08,SBC10} and highlight the regimes where the
quantum correlations and enhancement of transport coexist.

The paper is organized as follows. In Section\,\ref{Model} we define the discrete-time quantum walk model
used to describe the evolution process and transport efficiency. In
Section\,\ref{deco} the effects of depolarizing and dephasing noise on the transport process are discussed and in Section\,\ref{Qness} we use measurement induced disturbance to  highlight the
regime in which quantumness and efficient transport coexist. In  Section\,\ref{conc} we conclude.

\section{Model, evolution process and transport efficiency}
\label{Model}
Discrete-time quantum walk in one-dimension is modelled using the two internal degree of freedom of a particle on a position space.  The Hilbert space of the  particle  ${\cal H}_c$ is defined by, $|\uparrow \rangle = \begin{bmatrix} 1  \\ 0 \end{bmatrix}$ and $|\downarrow \rangle =\begin{bmatrix} 0  \\ 1 \end{bmatrix}$ as the basis state and the position Hilbert space ${\cal H}_p$ is defined by the basis state described in terms of $| j \rangle$, where $j \in {\mathbbm  I}$.  Each step of the walk comprises of the coin operation, $B (\theta)      \equiv  \begin{bmatrix} \begin{array}{clcr}
  \mbox{~~}\cos(\theta)      &     &     \sin(\theta)
  \\ -\sin(\theta) & &  \cos(\theta) 
\end{array} \end{bmatrix}$
which evolves the particle in superposition of its basis states followed by the  shift operation 
\be
S \equiv     \sum_j \left [  |\uparrow \rangle\langle
\uparrow|\otimes|j-1 \rangle\langle   j|   +  | \downarrow \rangle\langle
\downarrow|\otimes |j+1 \rangle\langle j| \right ]
\ee
which shift the particle in superposition of the position space. Therefore, the effective operation for each step of the discrete-time quantum walk is 
\be
\label{Wop}
W(\theta)\equiv S [B (\theta) \otimes  {\mathbbm 1}]
\ee
and the state after time $t$ ($t$ step), 
\be
|\Psi_t\rangle=W (\theta)^t|\Psi_{\rm in}\rangle,
\ee
where  
\be
|\Psi_{\rm in}\rangle= \left ( \cos(\delta/2)| \uparrow \rangle + e^{i\eta}
\sin(\delta/2)|\downarrow \rangle \right )\otimes | j=0\rangle.
\ee
The parameters $\delta$ and $\eta$ in the initial state and the parameter $\theta$ in the coin operation play a prominent role on the dynamics of walk.  In this paper we will restrict our study to walk with a symmetric transition of amplitude to its neighboring sites.  One of the configuration for the symmetric evolution of the walk will be to use a coin operation of the from $B = \frac{1}{\sqrt 2}\begin{bmatrix} \begin{array}{clcr}
 ~~ 1      &     &    -i
  \\ -i & &  ~~1 
\end{array} \end{bmatrix}$  for a particle with the initial state   $\frac{1}{\sqrt{2}}(|\downarrow\rangle 
                                       + |\uparrow\rangle)$.

\begin{figure}[H]
\bc
    \includegraphics[width=.23\linewidth]{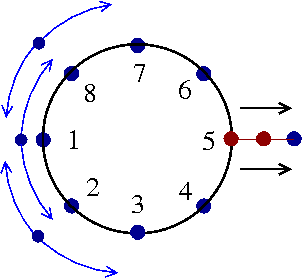}     
    \caption{(Color online) Schematic of a closed loop with $n=8$
      sites and sink potential at site $k=5$. Depending on the
      value of sink potential, a complete ($r=1$) or a fractional
      transfer ($r<1$) of the state takes place. \label{fig:Schematic} }
\ec
\end{figure}

Alternatively, for the symmetric evolution of the walk on an $n-$ cycle we can describe the dynamics by combining the coin and the shift operator. For this description, 
we will denote the wavefunction of the particle  as $|\Psi(t)\rangle=\sum_j| \psi(j,t)\rangle$, where the $\psi(j,t)$
represent two-component amplitude vectors of the particle at position
$j$ and time $t$,
\begin{equation}
  \label{compa0}
  |\psi(j,t)\rangle = (c_1|\downarrow \rangle + c_2|\uparrow \rangle) 
                      \otimes |j,t\rangle
                    = \begin{bmatrix}\psi_L(j,t) \\ \psi_R(j,t)\end{bmatrix}. 
\end{equation}
Here $L$ indicates the left-moving component and $R$ the right-moving
one.  The position space for the particle to move in is a closed loop
with $n$ discrete sites, $j=1,\dots,n$ (see  Fig.~\ref{fig:Schematic}), and we choose the wavefunction at $t=0$ 
to be localized at $j=1$ 
\begin{equation}
\label{ins}
  |\psi(1,0)\rangle = \frac{1}{\sqrt{2}}(|\downarrow\rangle 
                                       + |\uparrow\rangle)\otimes |1,0\rangle.
\end{equation}

After an evolution for time $t$ with equal left- and right-moving
components the particle's state will evolve into
\begin{equation}
  \label{eq:compa2}
  |\Psi(t)\rangle=\sum_j| \psi(j,t)\rangle 
                 =\sum_j\begin{bmatrix}\psi_L(j,t) \\ \psi_R(j,t) \end{bmatrix}
                 =W^t \begin{bmatrix} \psi_L(1,0) \\ \psi_R(1,0)  \end{bmatrix},
\end{equation}
where the evolution operator is given by
\begin{equation}
  W = \frac{1}{\sqrt{2}} 
  \left[\begin{tabular}{>{$}r<{$}>{$}r<{$}}
   a^{\dagger} & -ia^{\dagger} \\-ia &  a
        \end{tabular}\right].
\end{equation}
The operators $a$ and $a^{\dagger}$ are then defined by their effect on the wavefunction through
\begin{subequations}
  \begin{align}
    \label{eq:compa1}
    a \psi_{L (R)}(j, t)&= \psi_{L (R)}\Big((j+1)~\text{mod}(n+1), t+1 \Big),  \\ 
    a^{\dagger}\psi_{L (R)}(j, t)&= \psi_{L (R)}\Big ((j-1)~\text{mod}(n+1), t+1 \Big ),
  \end{align}
\end{subequations}
so that the evolution at every position $j$ can be calculated by using
\begin{subequations}
\begin{align}
\label{eq:compa}
  \psi_L(j,t+1)=\frac{1}{\sqrt{2}}
                \Big[&\psi_L\Big((j+1)\text{mod}(n+1),t\Big)
                  -i \psi_R\Big((j+1)\text{mod}(n+1),t\Big)\Big ],   \\
\label{eq:compb}
  \psi_R(j,t+1)=\frac{1}{\sqrt{2}}
                \Big[&\psi_R\Big((j-1)\text{mod}(n+1),t\Big) 
                  - i\psi_L\Big((j-1)\text{mod}(n+1),t\Big)\Big ].
\end{align}
\end{subequations}

To quantify the efficiency of the transport, we will designate one of the
position $k$ on the loop to act as a sink 
($k=5$ in Fig.\,\ref{fig:Schematic}). Depending on the strength of the sink
potential $r$ ($0 \leq r \leq 1$), a certain fraction of the wave
amplitude at this position is transported to a neighbouring loop/network. Therefore,  the
left- and the right-moving components
of the wavefunction at the neighbouring positions of the sink, $(k \pm 1)~\text{mod} (n+1)$,  at time $(t+1)$ after being absorbed at the sink are given by
\begin{subequations}
\label{eq:10}
\begin{align}
\label{eq:compak}
  \psi_L\Big((k + 1)~\text{mod}(n+1),t+1 \Big)
=\frac{1}{\sqrt{2}}
                \Big[\psi_L\Big((k+2)~\text{mod}(n+1),t\Big)
                  -i \sqrt{1-r}~\psi_R(k, t)\Big ],   \\
\label{eq:compb}
  \psi_R\Big((k + 1)~\text{mod}(n+1),t+1 \Big)=\frac{1}{\sqrt{2}}
                \Big[\sqrt{1-r}~\psi_R(k , t) 
                  - i\psi_L\Big((k+2)~\text{mod}(n+1),t\Big)\Big ], \\
  \psi_L\Big((k - 1) ~\text{mod}(n+1),t+1 \Big)=\frac{1}{\sqrt{2}}
                \Big[\sqrt{1-r}~\psi_L(k , t)
                  -i \psi_R\Big((k-2)~\text{mod}(n+1),t\Big)\Big ],   \\
\label{eq:compb1}
  \psi_R\Big((k - 1)~ \text{mod}(n+1),t+1 \Big)=\frac{1}{\sqrt{2}}
                \Big[\psi_R\Big((k-2)~\text{mod}(n+1),t\Big) 
                  - i\sqrt{1-r}~\psi_L(k , t)\Big ].
\end{align}
\end{subequations}
If the initial state of the particle at origin [Eq.\,(\ref{ins})] is written in the form of the density matrix,
\bea
\rho(0) = |\psi(1, 0)\rangle \langle \psi(1, 0)|,
\eea
the transport efficiency (TE) at time $t$ is given by
\begin{equation}
\label{teEQ}
  \text{TE} (t)= 1 - \sum_{j=1}^{n} \langle\psi(j, t)| \rho(t) |\psi(j, t) \rangle,
\end{equation} 
where the density matrix $\rho(t)$ is defined as
\begin{align}
 \rho(t) =& ({\mathbbm 1}_2\otimes s_k)W\rho(t-1)
              W^{\dagger}({\mathbbm 1}_2\otimes s_k)^\dagger, \nonumber \\
         =& S_k W \rho(t-1) W^\dagger S_k^\dagger.
\end{align}
Here $s_k$ is obtained by replacing 1 with $\sqrt{1-r}$ at position $(k, k)$
in an $n \times n$ unity matrix.  In Eq.\,(\ref{teEQ}), the $\sum_{j=1}^{n} \langle\psi(j, t)| \rho(t) |\psi(j, t) \rangle$ is the probability of finding the particle in the loop after some fraction being absorbed at the sink.

\begin{figure}[H]
\bc
\includegraphics[width=0.6\linewidth]{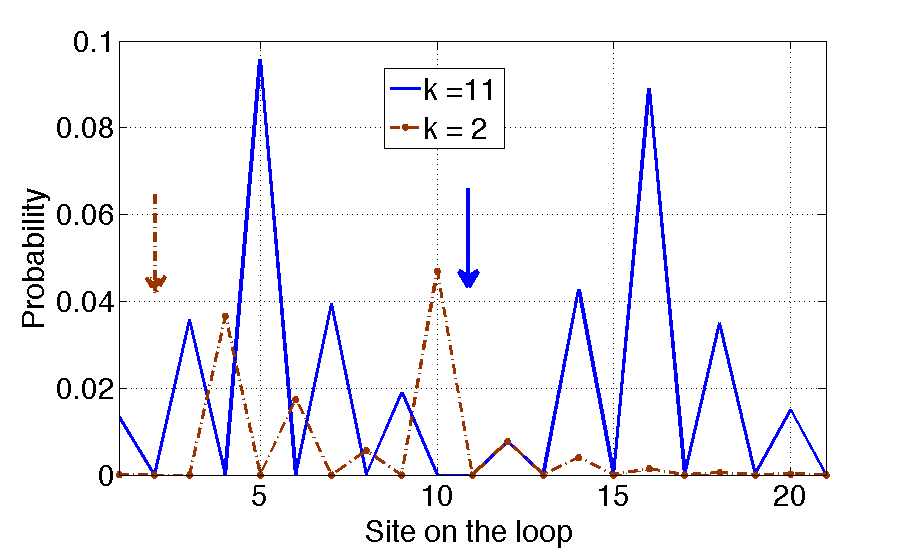}     
\caption{(Color online) Probability distribution in a
  loop of 21 sites at $t=25$ for a fully absorbing sink potential ($r=1$).  Shown are the two situations where the sink is located  the nearest to ($k=2$, red dashed line) or the farthest from ($k=11$, blue solid line)
  the initial position. 
The arrows indicate the position of the sink site.
\label{fig2}}
\ec
\end{figure}
\begin{figure}[H]
\begin{center}
  \subfigure[]{\includegraphics[width=7.8cm]{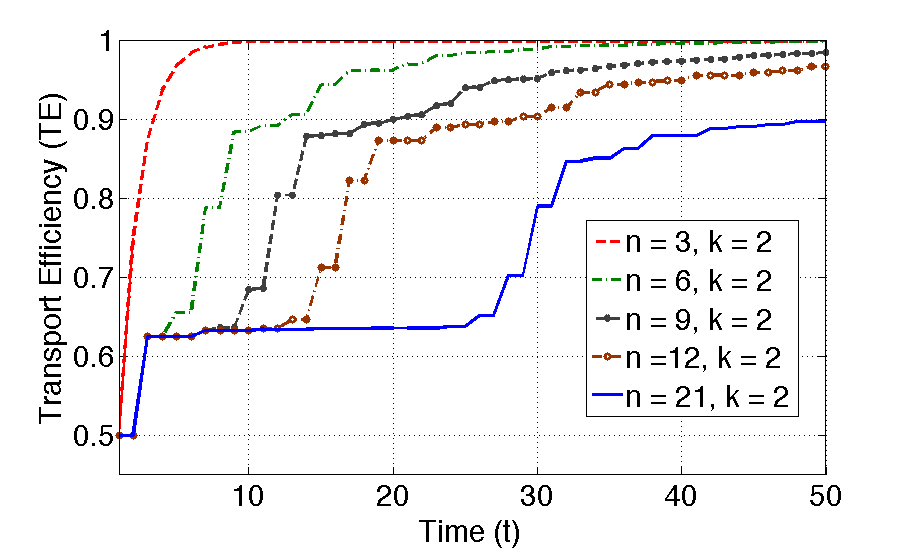}
    \label{fig:3a}} \hskip -0.15in
  \subfigure[]{\includegraphics[width=7.8cm]{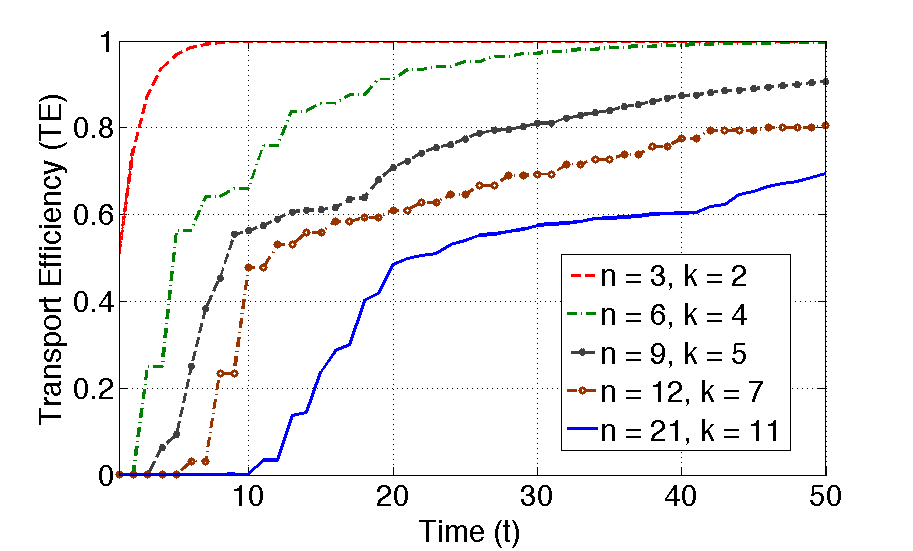}
    \label{fig:3b}}
\end{center}
\caption{(Color online) Transport efficiency as a function of time on
  a loop, when the sink
  potential $r=1$ is located at (a) the site nearest to the initial
  site and (b) furthest from the initial site. Note the different axis
  scalings.
\label{fig:3}}
\end{figure}

When sink potential has maximum strength ($r=1$), the right hand side
of the Eq.\,(\ref{eq:10}) consists only of a single term which describes 
complete absorption of the amplitude into the site $k$, which suppresses
the interference of the clockwise and counter-clockwise moving parts. In Fig.\,\ref{fig2} we show this situation for a loop of $21-$sites and the two situations in which the sink is either located at the site nearest ($k=2$) or the farthest away ($k=11$) from the initial position. In the first case half of the amplitude gets absorbed and transferred at $t=1$ and no longer
contributes to the interference on the ring for times $t>1$.  The
remaining  amplitude continues to evolve towards and away from $k$, which quickly leads to more of the amplitude being absorbed and transferred, especially 
if the number of sites in the loop is small (see Fig. \ref{fig:3a}). When $k=11$ the interference 
leads to the well known fast spreading of the amplitude around the loop, which  
results in an initially slower absorption rate, that even over time does not catch up with the one for the sink being at the nearest neighbour position (see Fig. \ref{fig:3b}).

\begin{figure}[ht]
\begin{center}
  \subfigure[]{\includegraphics[width=7.8cm]{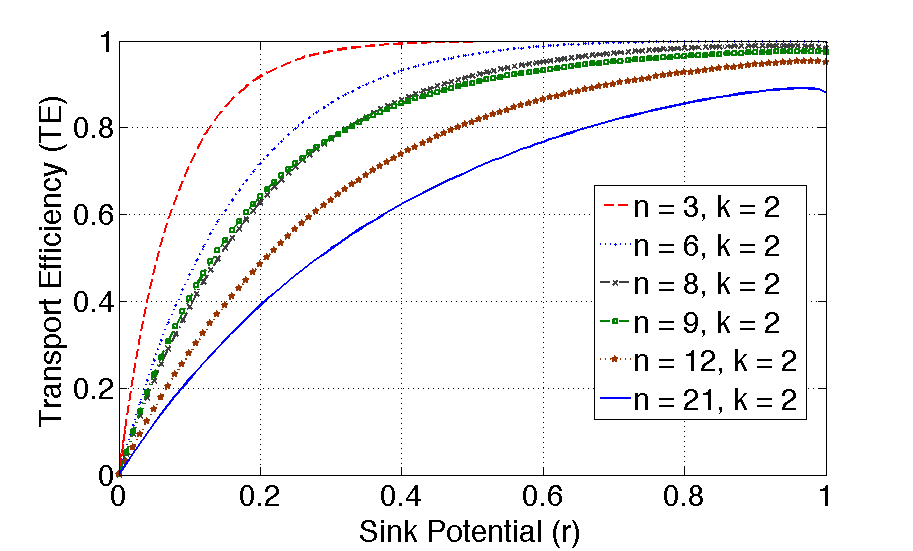}
    \label{fig:4a}} \hskip -0.15in
  \subfigure[]{\includegraphics[width=7.8cm]{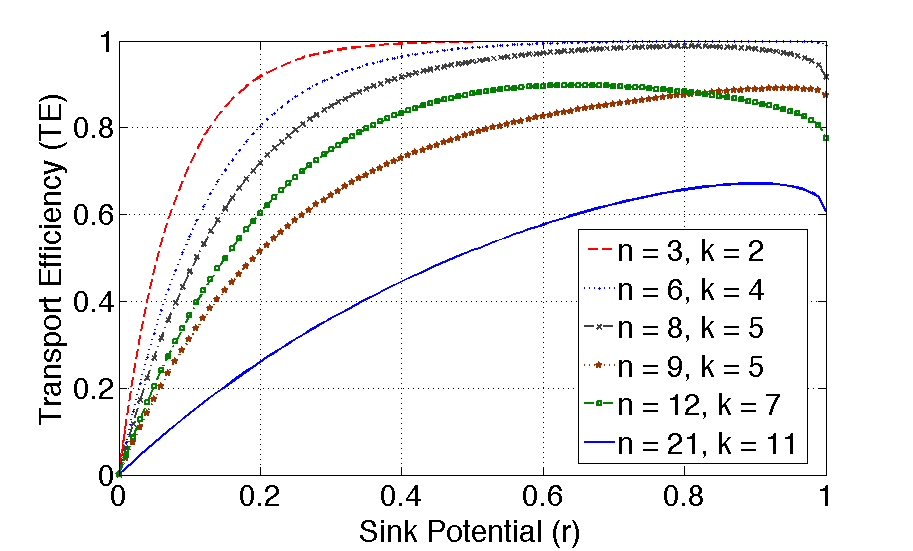}
    \label{fig:4b}}
\end{center}
\caption{(Color online) Transport efficiency as a function of sink potential $r$ at time $t=40$ (a)  when sink is the
  nearest neighbour ($k = 2$) of the initial site and (b) when sink is the
  farthest site from the initial site.
 \label{fig:4}}
\end{figure}

For $r<1$, a fraction of the amplitude ($\sqrt{1-r}$) at the position of the
sink is retained in the loop and continues to evolve and contribute to
the interference. When $k$ is the nearest neighbour, we still find a steep increase in the transport efficiency even for small values of $r$, which becomes less pronounced with increasing loop size (see Fig.\,\ref{fig:4a}). 
A similar behavior is found when $k$ is the farthest site from the initial position, however with increasing loop size, the curves also show  a local maximum for finite value of $r<1$ (see Fig.\,\ref{fig:4b}). This can be attributed to the role of quantum interference during crossover of the left and right moving components and is therefore a unique property of the ring geometry. 

\section{Decoherence}
\label{deco}

Decoherence can be introduced into the above model by considering a depolarizing and a dephasing
channel, which correspond to noise sources affecting the amplitude and
the phase of the particle. Their respective effects will be different and in this section we will discuss changes to the transport efficiency resulting from the presence of noise. 

\subsection{Depolarizing channel}

A depolarizing noise channel for a qubit system with density matrix $\rho_{2s}(t)$ can be described by replacing $\rho_{2s}(t)$ with a combination of a completely mixed state
and the unchanged state
\begin{align}
\tilde{\rho}_{2s}(t) = &\frac{P}{3}
           \left(\sigma_1\rho_{2s}(t)\sigma_1^\dagger+\sigma_2\rho_{2s}(t)\sigma_2^\dagger
                +\sigma_3\rho_{2s}(t)\sigma_3^\dagger\right)\nonumber \\
             &+(1-P)\rho(t)_{2s},
\end{align}
where $P\in[0,1]$ describes the strength of the noise and $\sigma_1$, $\sigma_2$, and $\sigma_3$ are the Pauli operators. For the discrete position space of the quantum walk to translates into 
\begin{align}
 \tilde{\rho}(t)= & S_k W \Big[(1-P)\tilde{\rho}(t-1)+\frac{P}{3}
           \Big({\mathbbm X}\tilde{\rho}(t-1){\mathbbm X}^\dagger
           \nonumber \\
             &+ {\mathbbm Y}\tilde{\rho}(t-1){\mathbbm Y}^\dagger
              +{\mathbbm Z}\tilde{\rho}(t-1){\mathbbm Z}^\dagger\Big)
          \Big] W^\dagger S_k^\dagger ,
\label{depol1}
\end{align}
with ${\mathbbm X} =\sigma_1 \otimes {\mathbbm 1}$, ${\mathbbm Y}
=\sigma_2 \otimes {\mathbbm 1}$ and ${\mathbbm Z} =\sigma_3 \otimes
{\mathbbm 1}$, from which the transport efficiency can be obtained as,
\begin{align}
TE (t) =  & 1 - \sum_{j =1}^{n} \langle \psi(j, t) | \tilde{\rho}(t)  |\psi(j, t)\rangle.
\label{depol}
\end{align}
\begin{figure}[H]
\bc 
\subfigure[ $n=7, k = 2$]{\includegraphics[width=6.3cm]{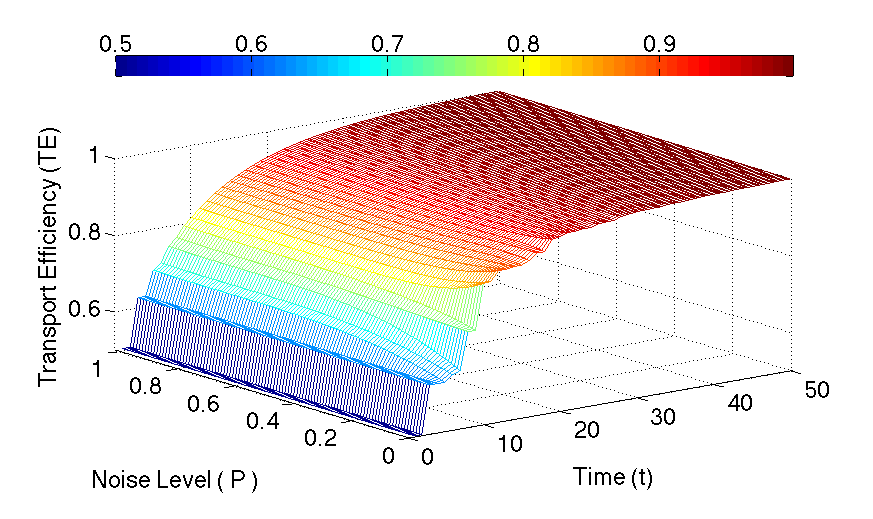} 
\label{fig:5a}}
\subfigure[ $n=7, k = 4$]{\includegraphics[width=6.3cm]{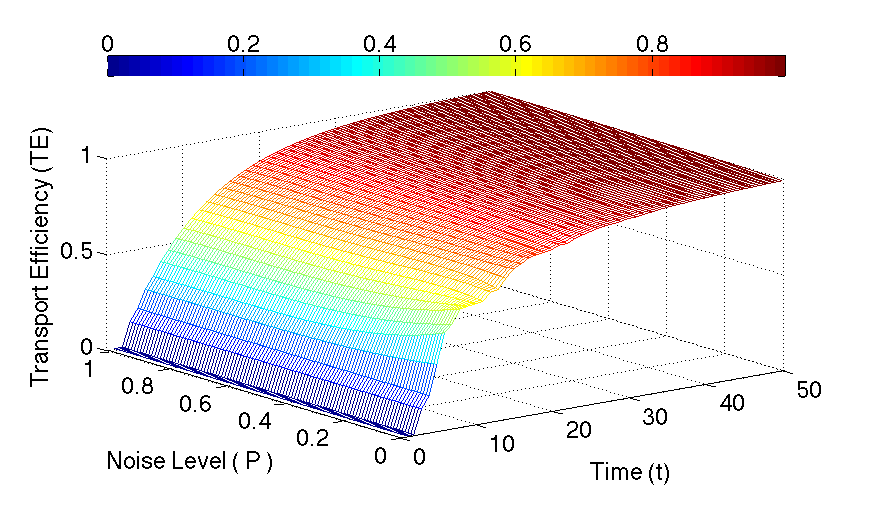}
\label{fig:5b}}\\
\subfigure[ $n=8, k = 2$]{\includegraphics[width=6.3cm]{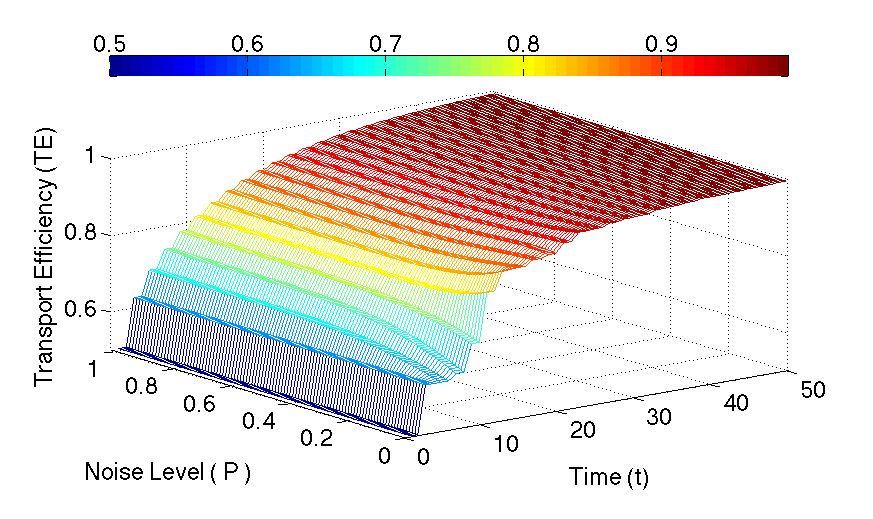}
\label{fig:5c}}
\subfigure[ $n=8, k = 5$]{\includegraphics[width=6.3cm]{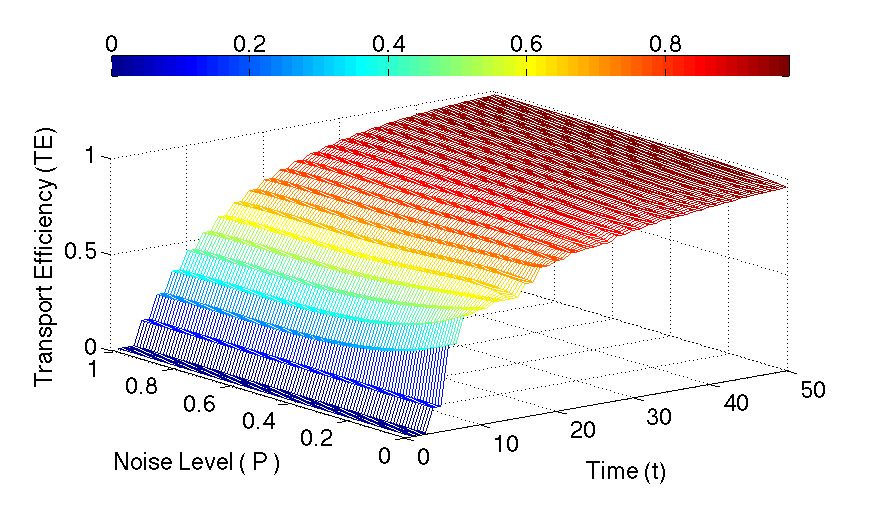}
\label{fig:5d}}\\
\subfigure[ $n=9, k = 2$]{\includegraphics[width=6.3cm]{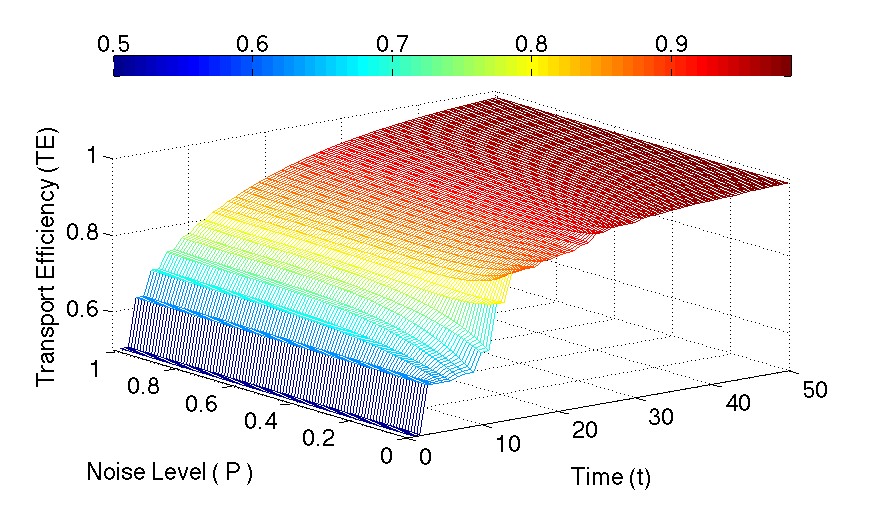} 
\label{fig:5e}}
\subfigure[ $n=9, k = 5$]{\includegraphics[width=6.3cm]{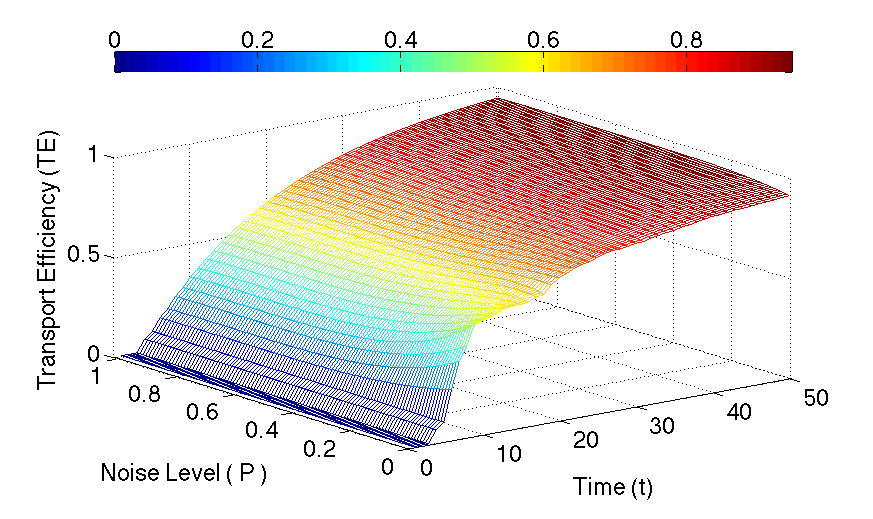}
\label{fig:5f}}\\
\subfigure[ $n=21, k = 2$]{\includegraphics[width=6.3cm]{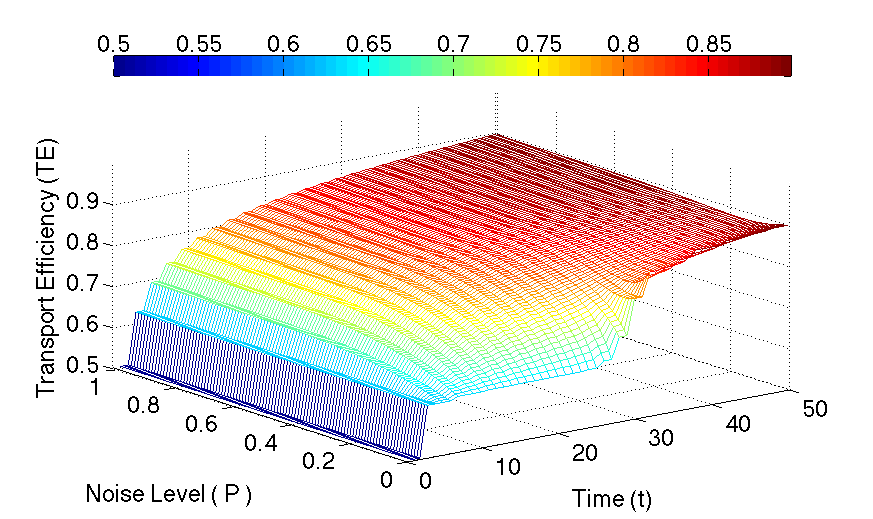}
\label{fig:5g}}
\subfigure[ $n=21, k = 11$]{\includegraphics[width=6.3cm]{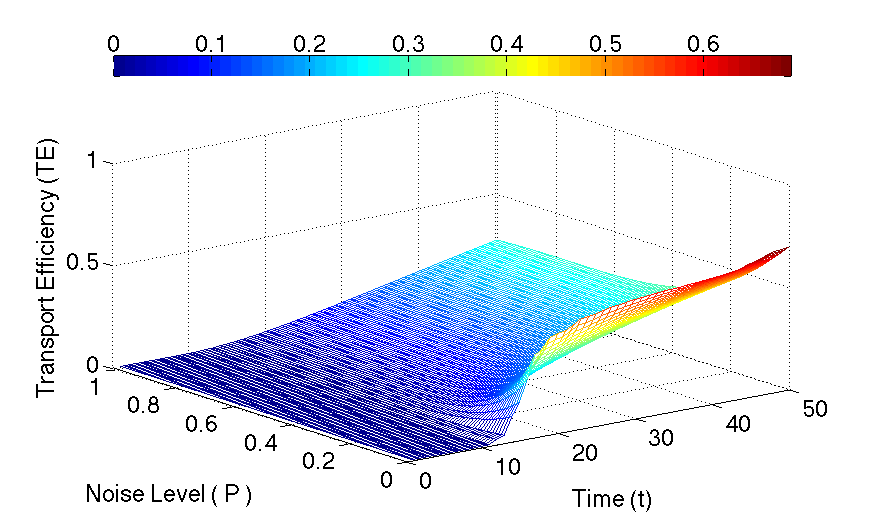}
\label{fig:5h}}\\
\subfigure[ $n=100, k = 2$]{\includegraphics[width=6.3cm]{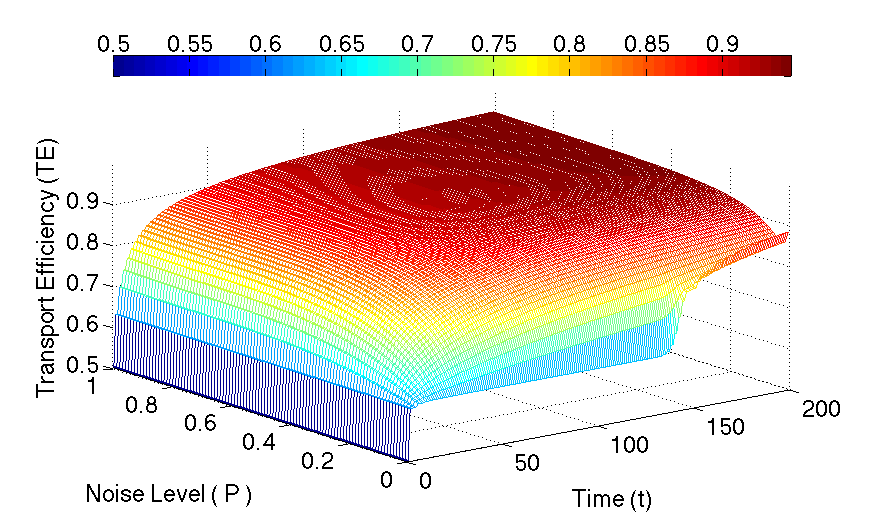}
\label{fig:5i}}
\subfigure[ $n=100, k = 51$]{\includegraphics[width=6.3cm]{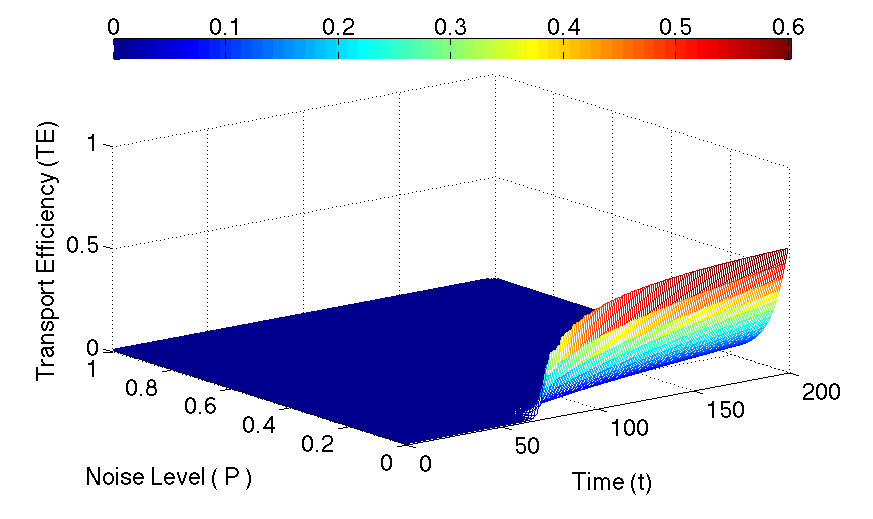}
\label{fig:5j}}
\ec
\caption{(Color online) Transport efficiency as a function of depolarizing
 noise level $P$ for a sink potential of $r=0.6$. In the left column the position of the sink is located on the nearest neighbour from the initial position  and in the right column on the farthest site. \label{fig:5}}
\end{figure}
\begin{figure}[H]
  \bc 
 \subfigure[ $n=8,k=2$]{\includegraphics[width=6.4cm]{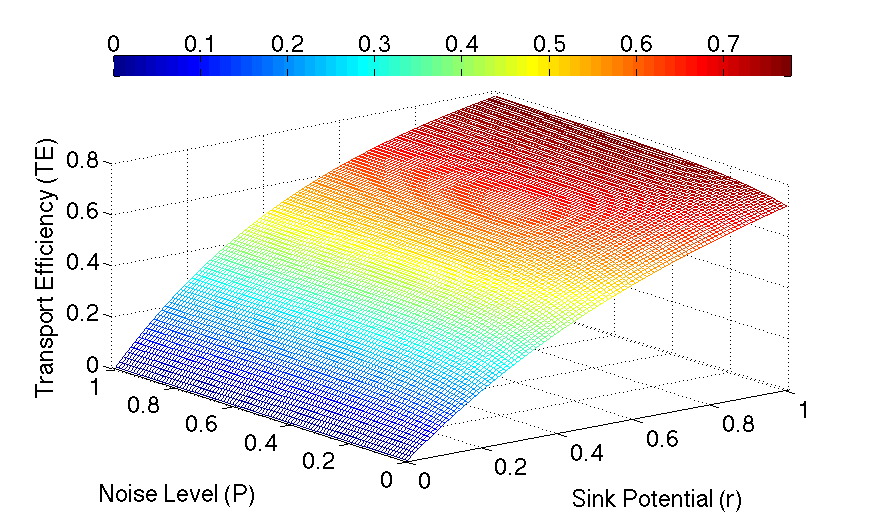}
 \label{fig:6a}}
  \hskip -0.1in 
 \subfigure[ $n=8,k=5$]{\includegraphics[width=6.4cm]{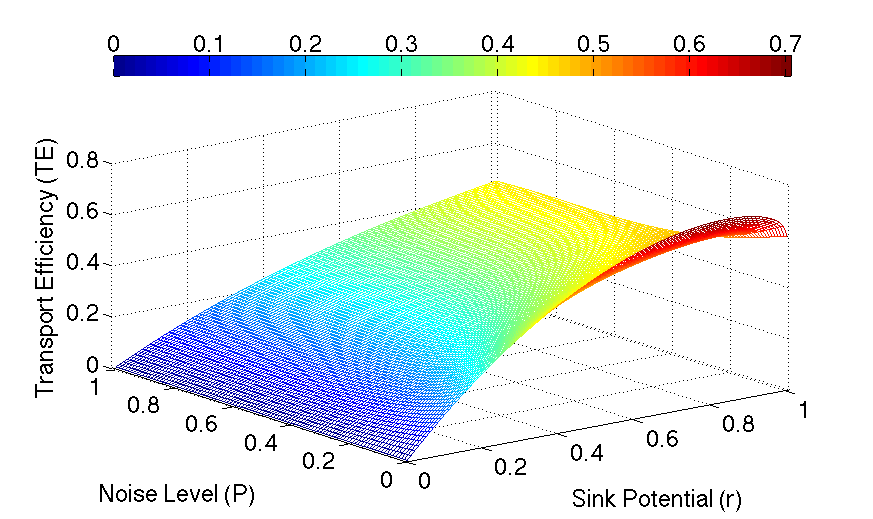}
    \label{fig:6b}}\\
\subfigure[ $n=9,k=2$]{\includegraphics[width=6.4cm]{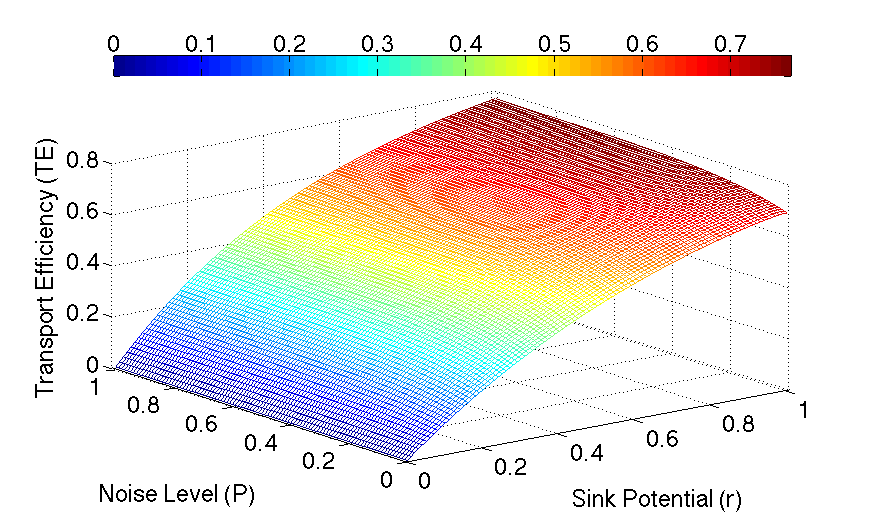}
 \label{fig:6c}}
  \hskip -0.1in 
 \subfigure[ $n=9,k=5$]{\includegraphics[width=6.4cm]{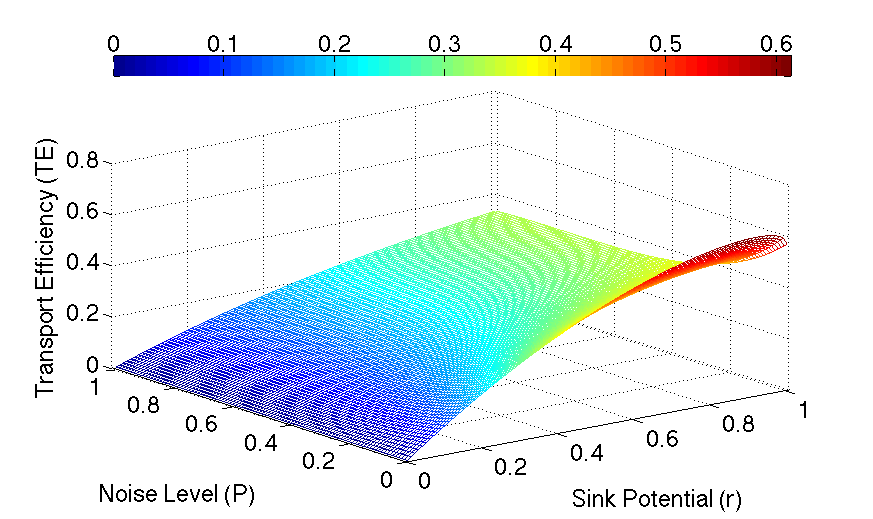}
    \label{fig:6d}}
  \ec
  \caption{(Color online) Transport efficiency as a function of depolarizing
 noise level $P$ and sink potential $r$ at $t=10$ on a closed loop of  $n=8$ and $n=9$ 
when the position of the sink ($k$) is the nearest neighbour and farthest site from the initial position.
 \label{fig:6}}
\end{figure}
\begin{figure}[t]
\bc 
\subfigure[ $n=7, k = 2$]{\includegraphics[width=6.4cm]{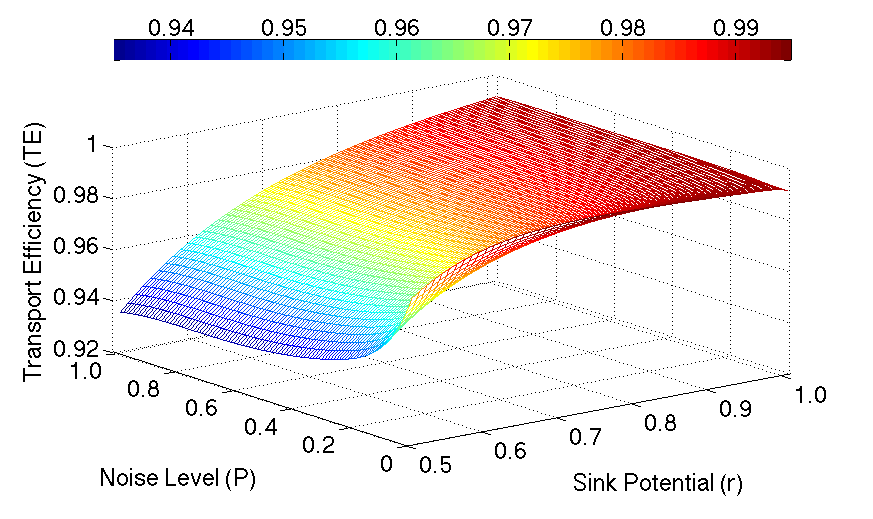} 
\label{fig:7a}}
\subfigure[ $n=7, k = 4$]{\includegraphics[width=6.4cm]{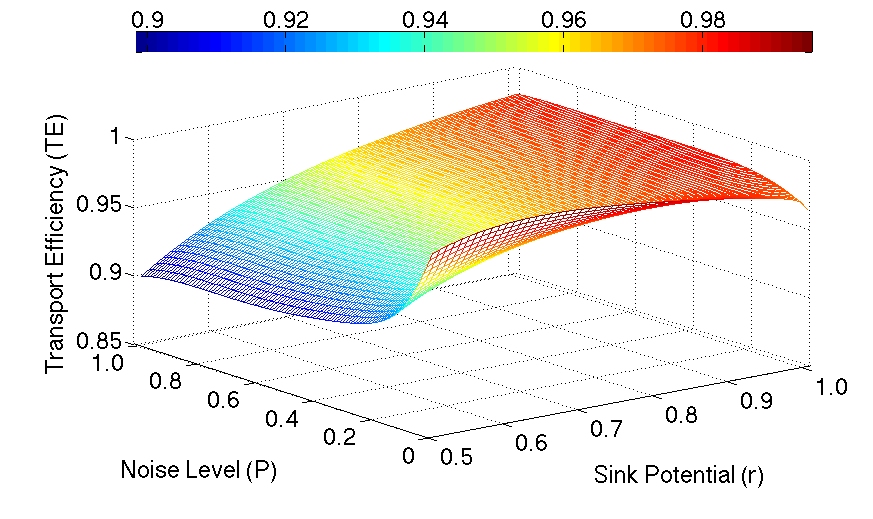}
\label{fig:7b}}\\
\subfigure[ $n=8, k = 2$]{\includegraphics[width=6.4cm]{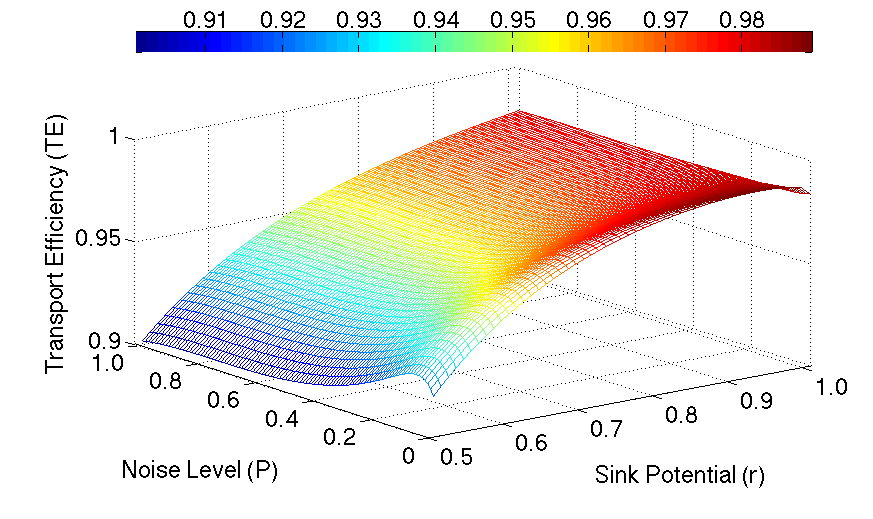}
\label{fig:7c}}
\subfigure[ $n=8, k = 5$]{\includegraphics[width=6.4cm]{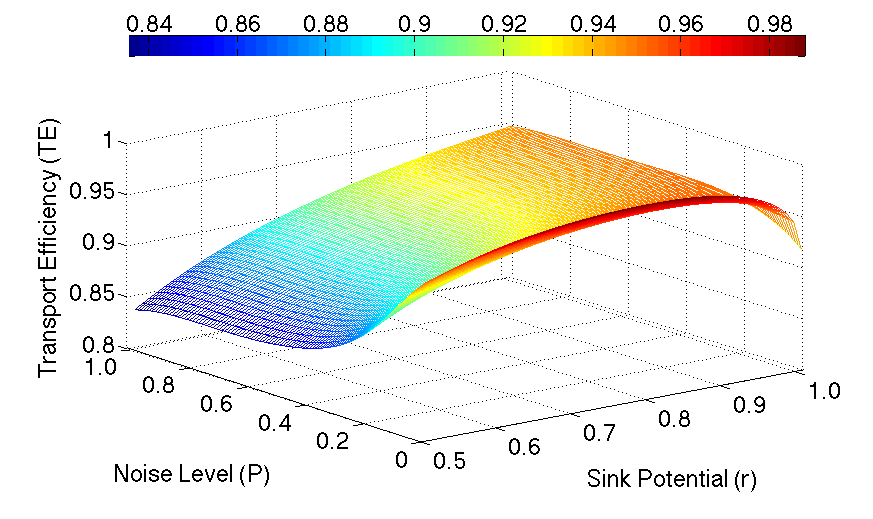}
\label{fig:7d}}\\
\subfigure[ $n=9, k = 2$]{\includegraphics[width=6.4cm]{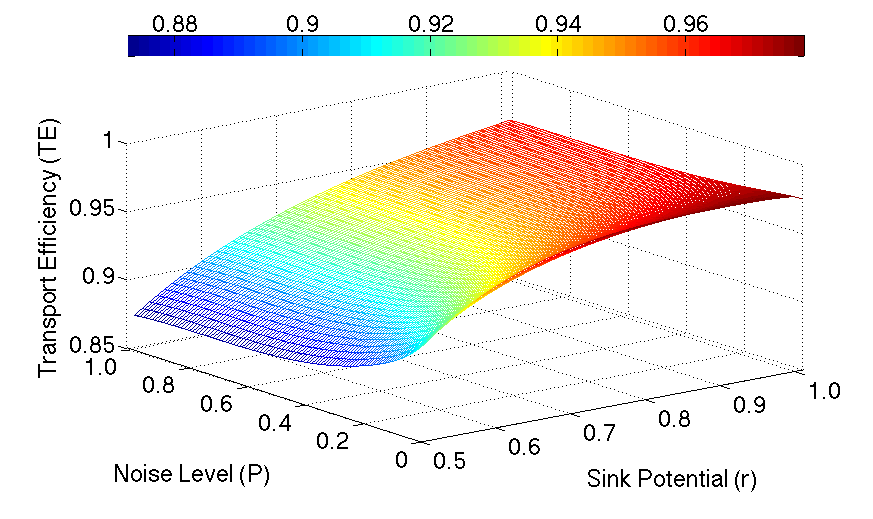}
\label{fig:7e}}
\subfigure[ $n=9, k = 5$]{\includegraphics[width=6.4cm]{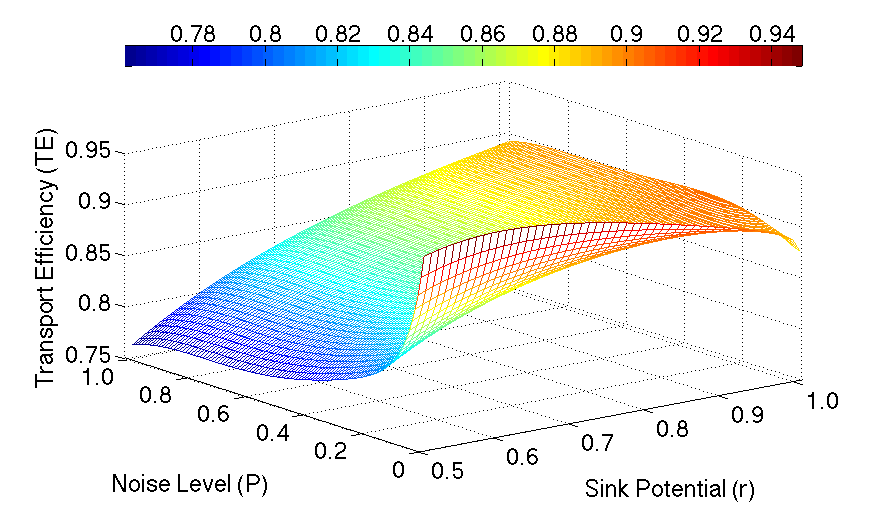}
\label{fig:7f}}\\
\subfigure[ $n=21, k = 2$]{\includegraphics[width=6.4cm]{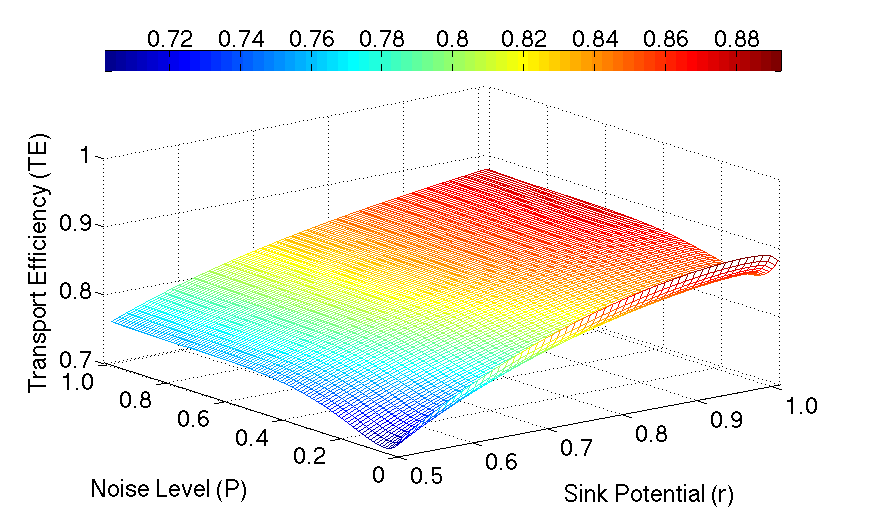}
\label{fig:7g}}
\subfigure[ $n=21, k = 11$]{\includegraphics[width=6.4cm]{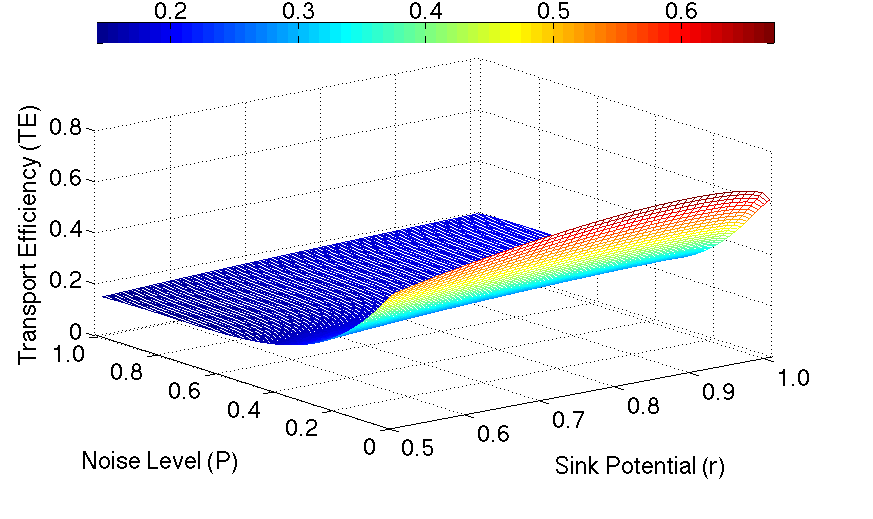}
\label{fig:7h}}
\ec
\caption{(Color online) Transport efficiency subject to a depolarizing
 noise channel on a closed loop of  different size ($n$) for the sink potential $r$ from $0.5$ to $1$ when the position of the sink ($k$) is the nearest neighbour and farthest site from the initial position at $t=40$. \label{fig:7}}
\end{figure}
The effect of this noise on transport efficiency when sink potential $r=0.6$ is shown in Fig.\,\ref{fig:5}. For short times and when the sink is the nearest site to the initial position  ($k=2$) one can see a clear increase in the  transport efficiency with increasing noise (see Fig.\,\ref{fig:5}, left column).  An inverse effect, that is a clear decrease in transport efficiency with noise is seen when the sink is the farthest site from the initial position  (see Fig.\,\ref{fig:5}, right column). 
A similar trend for all value of $r$  can be found at short times and we shown the representative situations for  $n=8$ and $n=9$  at $t=10$  in Fig.\,\ref{fig:6} .  An increase in transport efficiency can be seen when the sink is the nearest neighbour (Fig.\,\ref{fig:6}, left column) and a decrease when the sink is the farthest site (Fig.\,\ref{fig:6}, right column). For longer times the behaviour of the transport efficiency as a function of noise and sink potential shows a complex behaviour that does not immediately allow to identify general trends (see Fig.\,\ref{fig:7}). 
\begin{figure}[t]
\bc 
\subfigure[]{\includegraphics[width=6.4cm]{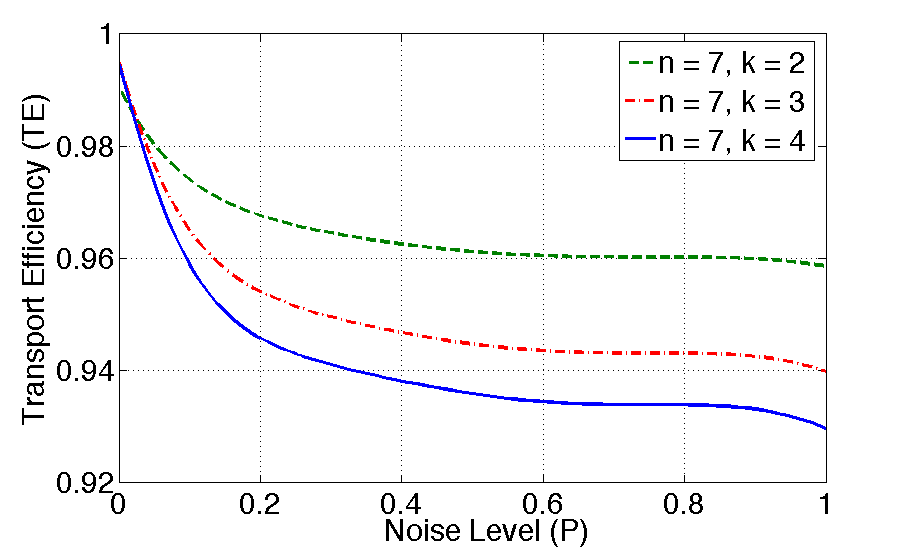} 
\label{fig:8a}}
\hskip -0.1in
\subfigure[]{\includegraphics[width=6.4cm]{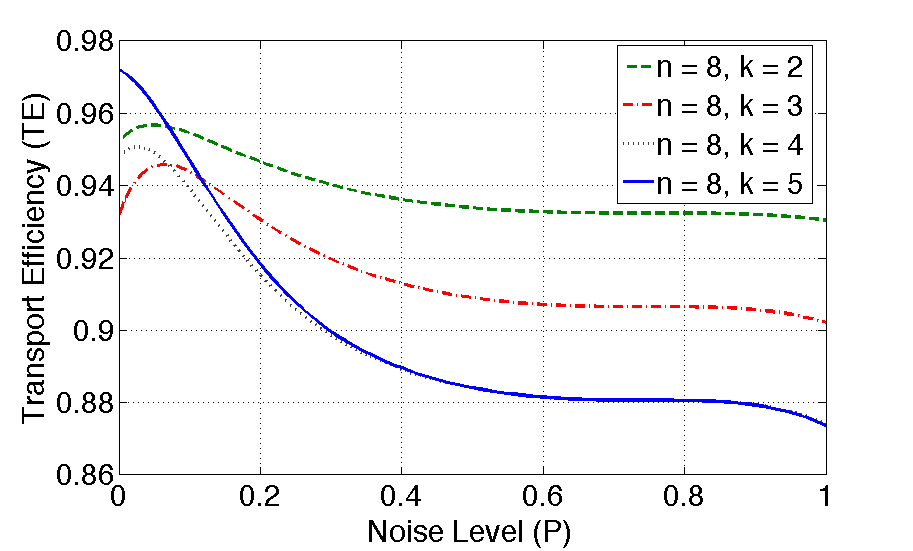}
\label{fig:8b}}\\
\subfigure[]{\includegraphics[width=6.4cm]{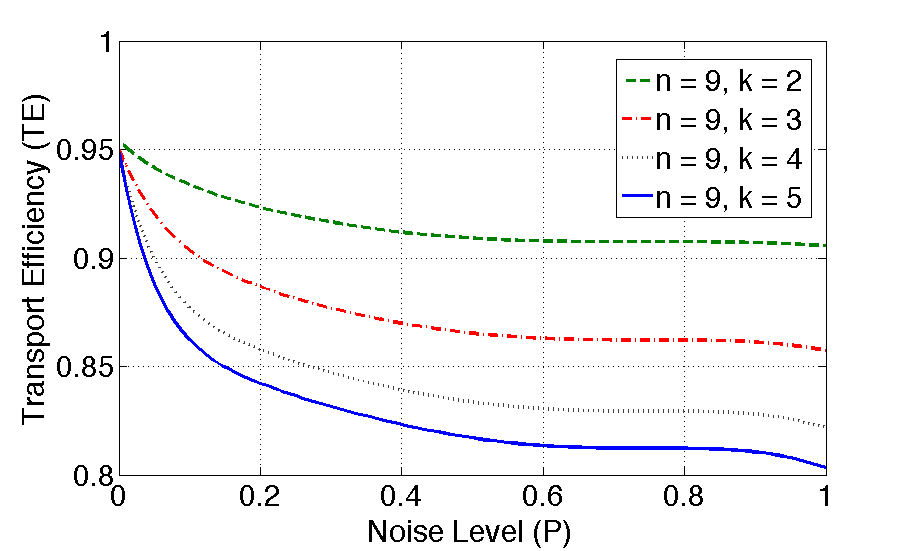}
\label{fig:8c}}
\hskip -0.1in
\subfigure[]{\includegraphics[width=6.4cm]{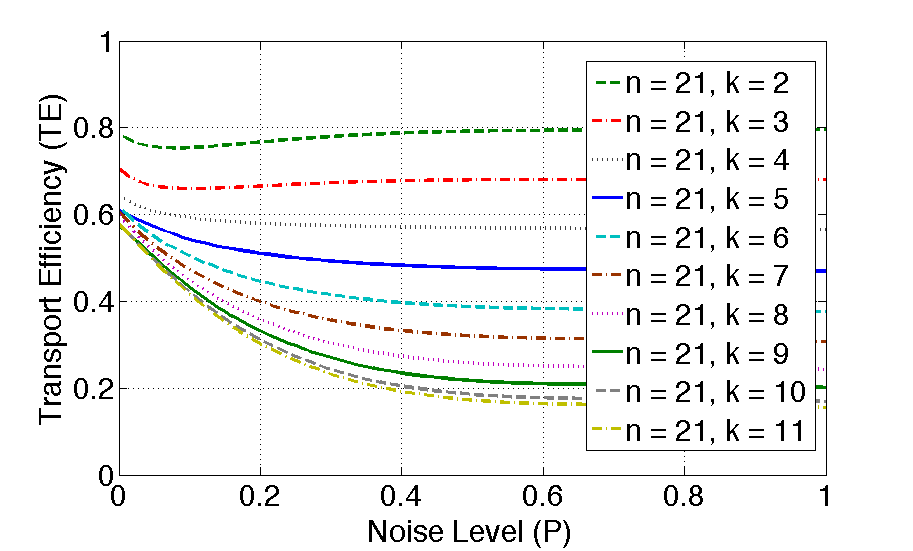}
\label{fig:8d}}
\ec
\caption{(Color online) Transport efficiency subject to a depolarizing
  noise channel on a closed loop of (a) 7 sites (b) 8 sites (c) 9
  sites and (d) 21 sites after $t = 40$ with $r=0.6$. 
  Note different axis scalings. \label{fig:8}}
\end{figure}
\begin{figure}[t]
\bc 
\subfigure[$n = 7$]{\includegraphics[width=6.5cm]{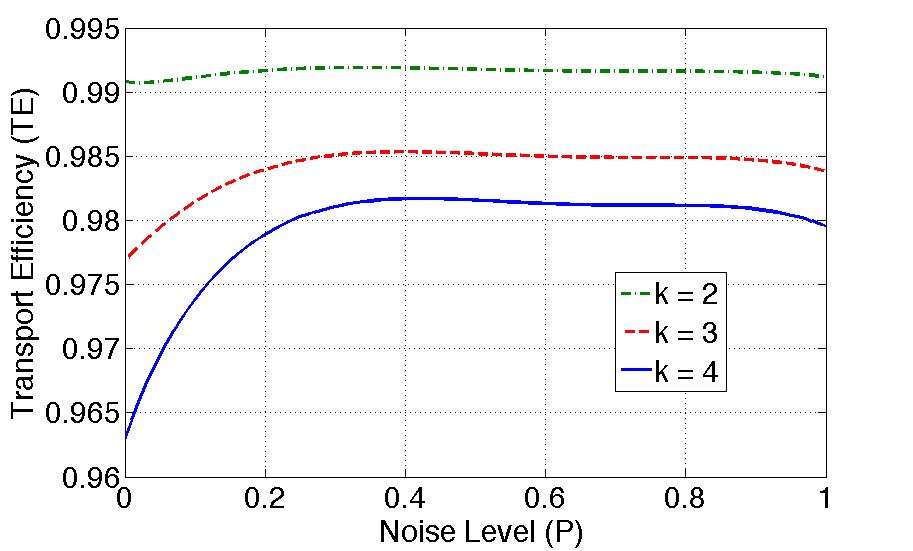} 
\label{fig:9a}}
\hskip -0.1in
\subfigure[$n=8$]{\includegraphics[width=6.5cm]{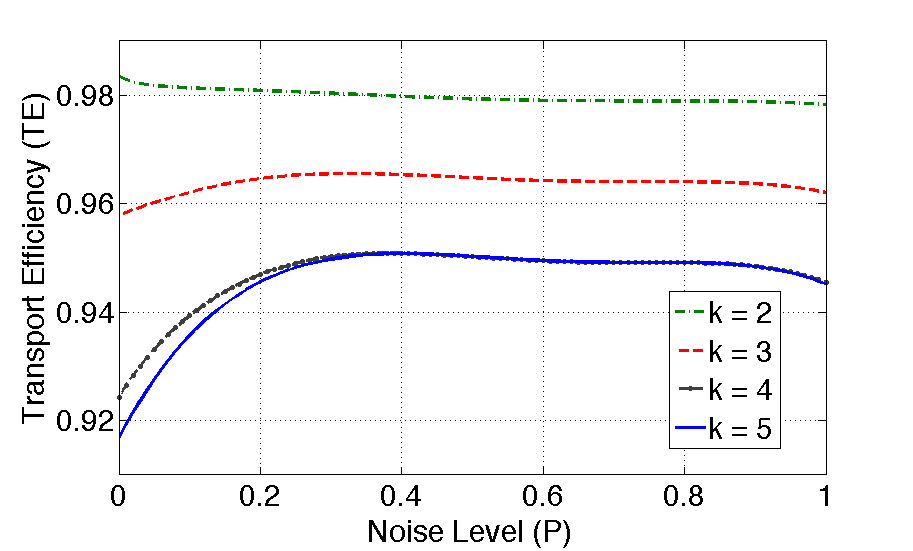}
\label{fig:9b}}\\
\subfigure[$n=9$]{\includegraphics[width=6.5cm]{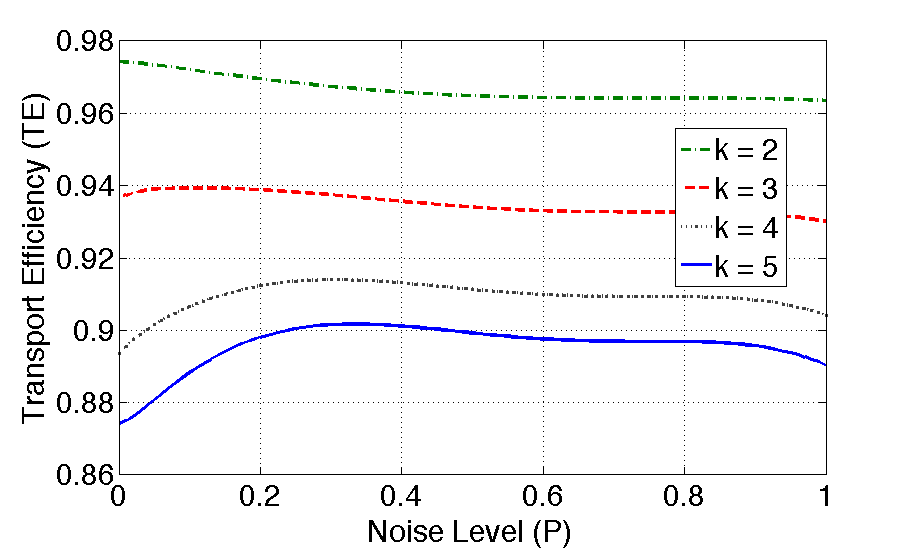}
  \label{fig:9c}} \hskip -0.1in
\subfigure[$n=21$]{\includegraphics[width=6.5cm]{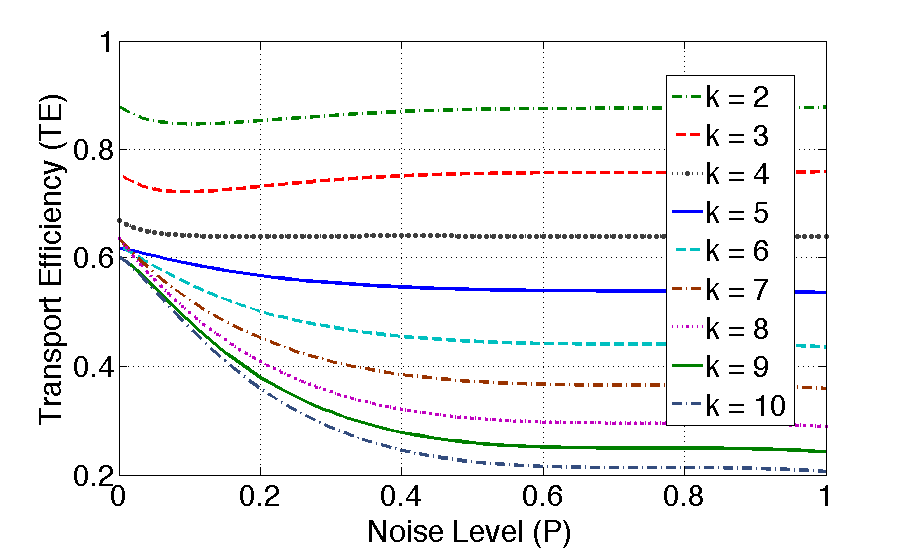}
\label{fig:9d}}
\ec
\caption{(Color online) Transport efficiency subject to a depolarizing
  noise channel on a closed loop of (a) 7 sites (b) 8 sites (c) 9
  sites and (d) 21 sites after $t = 40$ with $r=1$.  Note
  different axis scalings. \label{fig:9}}
\end{figure}
\begin{figure}[t]
\bc
\subfigure[8-site with sink, k = 2]{\includegraphics[width=6.5cm]{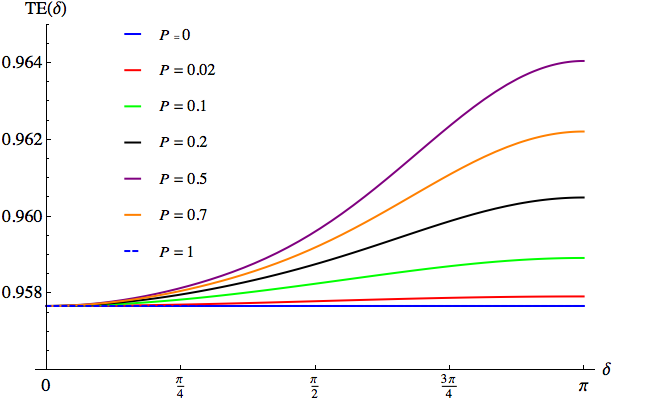} 
\label{fig:10a}}
\hskip -0.1in
\subfigure[8-site with sink, k= 5]{\includegraphics[width=6.5cm]{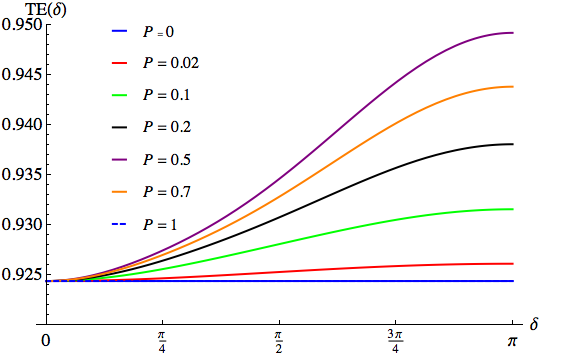}
\label{fig:10b}}
\ec
\caption{(Color online) Transport efficiency with sink at
  position (a) $2$ and (b) $5$ on an $8-$site ring as a function of
  $\delta$ for different dephasing noise level at $t=40$. For any
  non-zero noise level transport efficiency is maximum for $\delta =
  \pi$.
  \label{fig:10}}
\end{figure}
\begin{figure}[t]
\bc 
\subfigure[8-site with sink, k= 2]{\includegraphics[width=6.5cm]{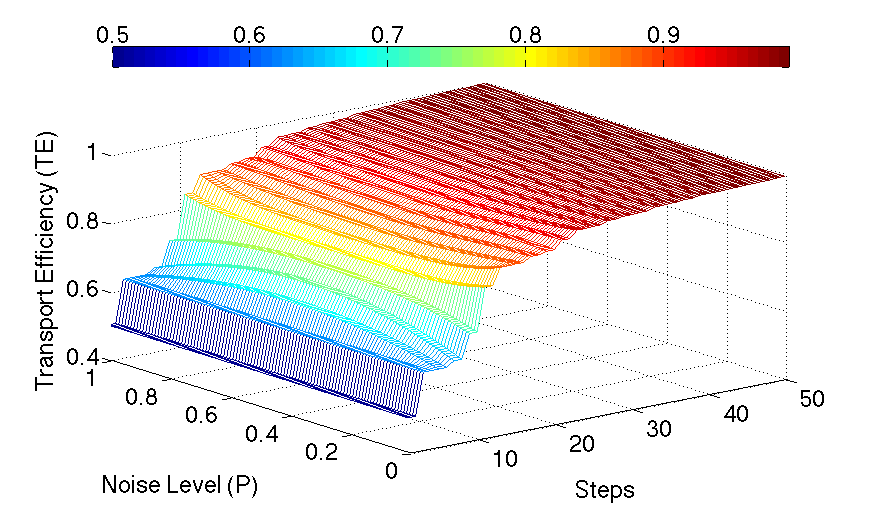} 
\label{fig:11a}}
\hskip -0.1in
\subfigure[8-site with sink, k = 5]{\includegraphics[width=6.5cm]{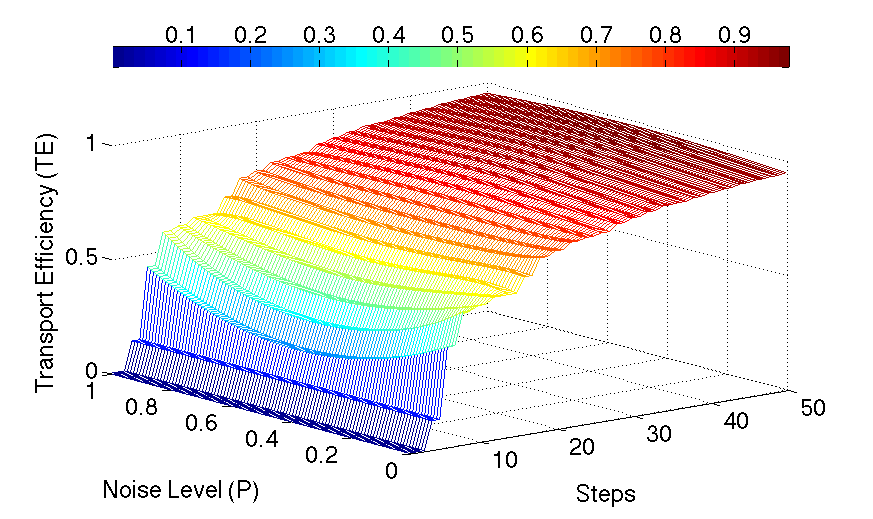}
\label{fig:11b}}
\ec
\caption{(Color online) Transport efficiency as a function of  dephasing noise level $P$ and $t$ (steps) with $\delta = \pi$ and $r=1$ on a $8-$site ring with the sink, $k=2$ and $k=5$. \label{fig:11}}
\end{figure}
\begin{figure}[t]
\bc 
\subfigure[8-site with sink, k = 2]{\includegraphics[width=6.5cm]{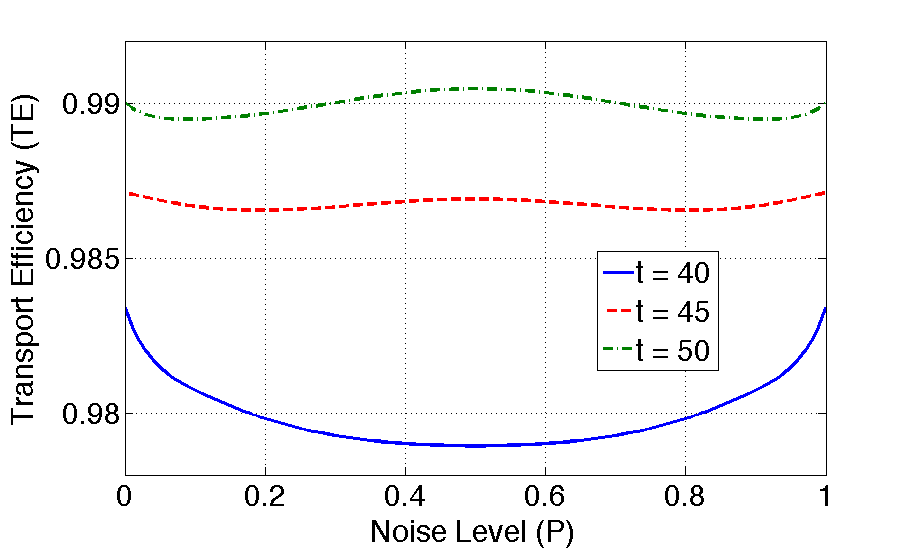} 
\label{fig:12a}}
\hskip -0.1in
\subfigure[8-site with sink, k = 5]{\includegraphics[width=6.5cm]{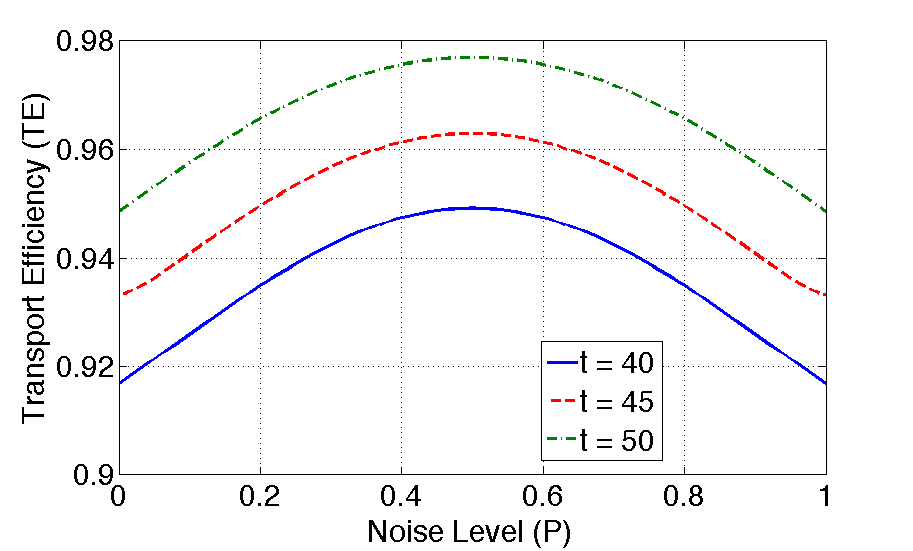}
\label{fig:12b}}
\ec
\caption{(Color online) Transport efficiency of two-state particle
  with dephasing noise with $\delta = \pi$ on a $8-$site ring with the sink, (a) $k=2$ and (b) $k=5$ 
at $t=40$, $t=45$, and $t=50$. \label{fig:12}}
\end{figure}
\begin{figure}[t]
\bc 
\subfigure[7-site with sink, k = 2]{\includegraphics[width=6.5cm]{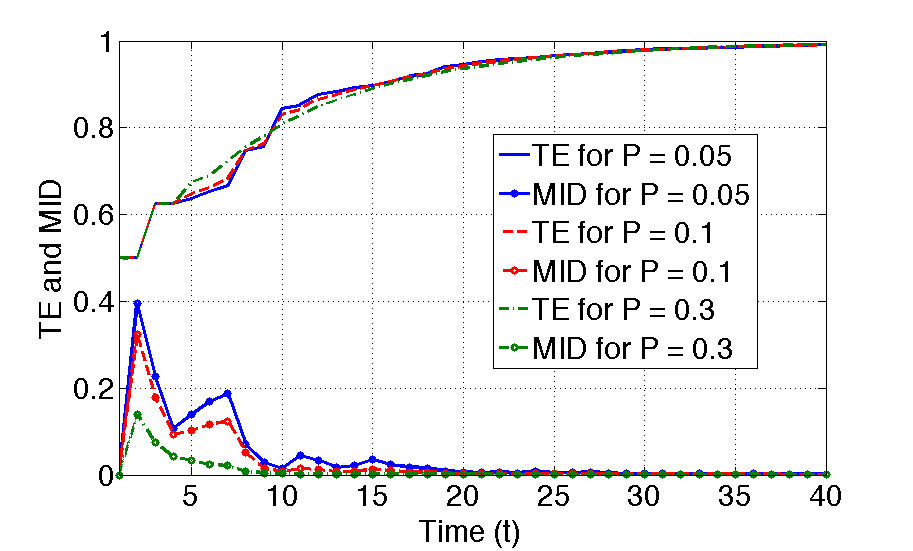} 
\label{fig:13a}}
\hskip -0.1in
\subfigure[7-site with sink, k = 4]{\includegraphics[width=6.5cm]{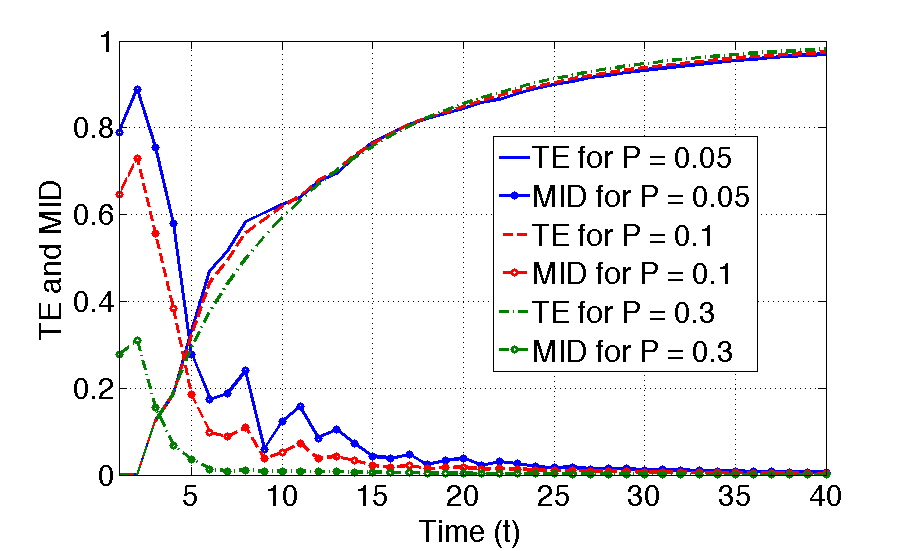}
\label{fig:13b}}\\
\subfigure[8-site with sink, k = 5]{\includegraphics[width=6.5cm]{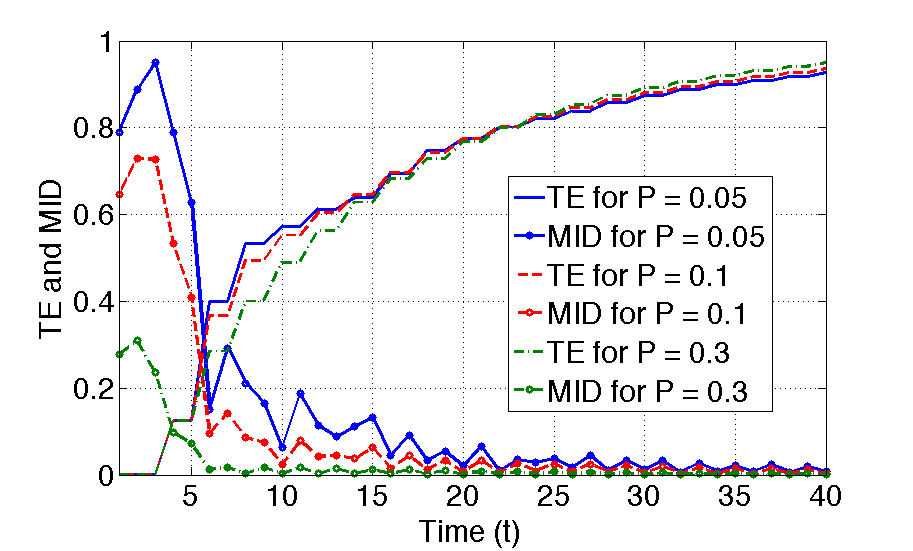}
\label{fig:13c}}
\hskip -0.1in
\subfigure[9-site with sink, k = 5]{\includegraphics[width=6.5cm]{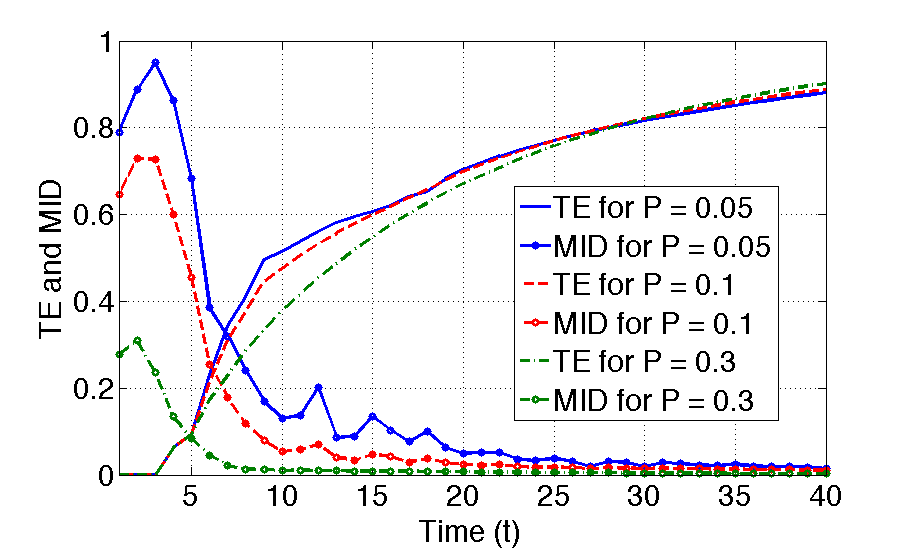}
\label{fig:13d}}
\ec
\caption{(Color online) Transport efficiency of two-state particle and
  quantum correlations between the particle and the ring when
  subjected to the decoherence due to depolarizing channel with time $t$ when $r=1$ for loop with (a)
$n=7$, $k=2$ (b) $n=7$, $k=4$ (c) $n=8$, $k=5$ and (d) $n=9$, $k=5$. \label{fig:13}}
\end{figure}
\par
To illustrate this complex behaviour more clearly we show in Figs.\,\ref{fig:8} and \ref{fig:9} the transport efficiency as a function of noise strength for two different value of $r$ at $t=40$ with the sink placed at different positions. For $r=0.6$ we find a general decrease in transport efficiency with increasing $P$ and distance of sink from the initial position (Fig.\ref{fig:8}). A noteworthy exception from this trend can be seen for $n=8$, where a small increase in transport efficiency is found for very small levels of noise. However, with increasing noise this quantity follows the general trend again.

For $r=1$ and the sink site $k$ being far away from the initial position we find a clear increase in transport efficiency for values up to $P\approx 0.3$ (see Figs\,\ref{fig:9}(a-c)), however with increasing loop size and distance between the initial position and sink, a steep decrease in the transport efficiency 
is seen for all $r$ (see examples in Fig.\ref{fig:8d} and Fig.\ref{fig:9d}). This is also very clearly visible from Fig.\,\ref{fig:7h}.
\par
The reason for this complex behaviour can be found in the fact that the noise changes the ballistic-like probability distribution of the particle over time into a  Gaussian-like distribution around the initial position \cite{CSB07}.  For short times ($t=10$) and when sink is the nearest neighbour to the initial position, this increases the fraction of the amplitude staying close to the initial position and consequently increases the transport efficiency. If the sink is far away from the initial position and the loop has only a small number of sites ($n\approx7,8,9$), the fraction of the amplitude that reaches the sink is smaller in presence of noise, leading to the described decrease. 

For longer times and $r<1$,  the crossover of the clock-wise and counter clock-wise moving part of the amplitude at the sink result in the mixing of the amplitude and wider spread of the probability distribution on the loop. In the presence of noise this spreading decreases, which results in a decrease of amplitude moving towards sink, which in turn affects the transport efficiency. Therefore, with an increase in distance of the sink from the initial position an increase in the degradation of transport efficiency is seen (see Fig.\,\ref{fig:8} for $r=0.6$). When $n$ is small and the sink away from the initial position, the small increase in transport efficiency with noise for $r=1$ is mainly due to the absence of the mixing of the clockwise and counter clockwise moving component of the amplitude.

\subsection{Dephasing channel}

The density matrix of a two-state quantum walker in the presence of a dephasing channel can be written as
\begin{align}
  \tilde{\rho}(t) = S_k W \left[P{\mathbbm E}\tilde{\rho}(t-1){\mathbbm E}^\dagger 
                        + (1-P)\tilde{\rho}(t-1)\right] W^\dagger S_k^\dagger,
\label{dephas}
\end{align}
 where
\begin{align}
 {\mathbbm E} = |\uparrow \rangle\langle\uparrow|+e^{- i \delta}
                |\downarrow\rangle\langle\downarrow|,
\end{align}
and $P$ and $\delta$ are the dephasing noise level and angle,
respectively. For $\delta = 0$, ${\mathbbm E}$ is identity corresponding to a
noiseless evolution and for $\delta = \pi$ it represents a phase
flip channel. When $P=1$ and $\delta = \pi$  the evolution in Eq.(\ref{dephas}) is identical to a phase flip operation at each step of the quantum walk and the resulting probability distribution is identical to the noiseless evolution. This can be seen as a symmetry of the quantum walk \cite{CSB07} and indicates that using a dephasing noise channel a maximally mixed state $\tilde{\rho}(t)$ is obtained for $P = 0.5$. Note that no such symmetry exists in the presence of depolarizing noise, which means that a maximally mixed state $\tilde{\rho}(t)$ in Eq.(\ref{depol1}) is obtained only for $P=1$.

In Fig.~\ref{fig:10} we show the transport efficiency as a function of
$\delta$ for different noise levels $P$ for an $8-$site loop with the
sinks at $k=2$ and $k=5$ at time $t=40$ when the sink potential is
$r=1$.  One can clearly see an increase in the transport efficiency in
the presence of depolarizing noise, which gets more pronounced the
larger the noise and the dephasing angle. However, the difference
between the maximum and the minimum of the transport efficiency is
very small and given by $0.6$\% when the trap site is the nearest
neighbour and $2.5$\% when the trap site is the farthest from the
initial position.  

The transport efficiency  for fixed $\delta = \pi$ as a function of 
$t$ and noise level $P$ is shown in Fig.~\ref{fig:11} . For short times ($t=10$) and when the sink is the nearest neighbour to the initial position, one can clearly see an increase for any value of $P$, with the maximum located at $P=0.5$ \,(see Fig.\,\ref{fig:11a}).  Conversely, for the sink being farthest away from the initial position ($k=5$ for $n=8$), a suppression of the transport efficiency is found, with the minimum located at $P=0.5$ \,(see Fig.\,\ref{fig:11b}).  For long times, the effect of noise on the transport efficiency becomes too small to be visible (see Fig.\,\ref{fig:11}) and we show in Fig.\,\ref{fig:12} that the detailed trend is inverse to the one seen for short times.

The effects of a dephasing noise channel are therefore very similar to
the effects stemming from a depolarizing noise channel discussed in the
preceding section. However, there is one major difference in that the extrema for the transport efficiency
for the dephasing channel are achieved at $P=0.5$\,\cite{CSB07}. 

\section{Quantumness in the transport process}
\label{Qness}

Let us finally discuss the quantum nature of the above walks in order to identify the role quantum correlations play in the appearance of 
transport enhancements. To quantify this we will make use of the
concept of measurement induced disturbance (MID)\,\cite{L08}, which can be seen as a measure of quantum correlations between the closed loop with the sink and the surviving part of the particle state as the subsystems. 

MID assumes that the quantum correlations of the bipartite state $\tilde{\rho}$ living in the
Hilbert space ${\cal H}_A \otimes {\cal H}_B$ can be obtained by using
a reasonable measure of total correlations and subtracting a
reasonable measure of classical correlations.  If
the reduced density matrices are denoted by $\tilde{\rho}_A$ and $\tilde{\rho}_B$, a
reasonable measure of total correlations between systems $A$ (loop) and $B$ (particle)
is the mutual information, given by
\begin{equation}
  I(\tilde{\rho}) = S(\tilde{\rho}_A) + S(\tilde{\rho}_B) - S(\tilde{\rho}),
  \label{eq:I}
\end{equation}
where $S(\tilde{\rho}) = - \mathrm{Tr}(\tilde{\rho} \ln \tilde{\rho})$ denotes the von Neumann
entropy.  By taking the complete projective measurement 
$\Pi_A \otimes \Pi_B$ determined by the eigen-projectors $\Pi_A^i$  and 
$\Pi_B^i$ of the marginal states $\rho_A = \sum_j p_A^i\Pi_A^i$ and $\rho_B = \sum_j
p_B^i\Pi_B^i$, where $p_A$ and $p_B$ are the corresponding eigenvalues, one can define the 
MID as
\begin{equation}
  \label{eq:Q}
  Q(\tilde{\rho}) = I(\tilde{\rho}) - I[\Pi(\tilde{\rho})].
\end{equation}
In the preceding expression 
\begin{equation}
  \label{eq:Pi}
  \Pi(\tilde{\rho}) \equiv \sum_{j,k} \Pi_A^j \otimes \Pi_B^k \tilde{\rho}
  \Pi_A^j \otimes \Pi_B^k
\end{equation}
is the post-measurement state after local measurements $\Pi_A$ and $\Pi_B$. 
The state $\Pi(\tilde{\rho})$ may be considered classical in the sense that
there is a (unique) local measurement strategy, namely $\Pi$, 
that leaves $\Pi(\tilde{\rho})$ unchanged. This strategy is special in that it produces a 
classical state in $\rho$ while keeping the reduced states invariant.  If we accept that
$I[\Pi(\tilde{\rho})]$ is a good measure of {\it classical} correlations in
$\tilde{\rho}$, then one can consider MID as a computable measure of quantum correlation \cite{L08}.

In absence of noise, quantum correlations between the loop and the particle 
vanish only when the particle is completely transported to the sink, that is, the transport
efficiency is one (TE=1).  In the presence of noise, however, we find that
quantum correlations vanish before the state is completely
transported to the sink. In Fig.~\ref{fig:13}, we show the transport
efficiency and the corresponding quantum correlations as a
function of time for different depolarizing noise levels. One can see
that even a small increase in the noise level leads to a significant
decrease in the strength of the quantum correlations and they vanish
even before the transport efficiency gets close to one. This shows that the quantum
correlations are only present for a very short amount of time during
the transport process.  From Fig.~\ref{fig:13}, we can note that the quantum correlations
during the transport process on a loop of small size (7, 8, 9) reduce faster with time when the 
sink is the nearest neighbour than when the sink is the farthest site. From this one can draw the conclusion that an enhancement of transport efficiency can be seen when the noise level is very small, the sink is the nearest neighbour to the initial position, and the evolution has been brief. In this regime quantum correlations and
enhancement of transport efficiency coexist.

Let us finally point out that if the sink is connected to a second loop and if the quantum correlations between the surviving particle and the loop are non-zero, then non-zero quantum correlation between the second loop and the initial loop will be established.

\section{Conclusion} 
\label{conc}
Using a discrete-time quantum walk as a transport process we have
studied the transport efficiency of a quantum state on a closed
loop. In particular, we looked at the enhancement of quantum
transport in presence of a depolarizing and a
dephasing noise channel.  We have shown that these noise channels can enhance
the transport efficiency only when the number of sites in the closed
loop is small. In addition, we have found enhancement for short times
when the sink is the nearest neighbour of the initial site and for
longer times $t$, when the sink is farthest from the initial
position. 

Furthermore we investigated the loss of quantum
correlations due to the presence of noise and  identified
regimes where noise-enhanced transport and significant quantum
correlations coexist. Such regimes are important for example in systems in which energy or electrons are transferred from one loop
or molecule to another, as they allow to generate quantum correlations
between the different systems. Together with helping to understand and model the energy transfer and state transport processes in naturally occurring system, the recent
experimental progress in creating quantum walks in various physical
systems (NMR\,\cite{DLX03,RLB05,LZZ10}, cold ions\,\cite{SMS09,ZKG10}, photons\,\cite{PLP08,PLM10,SCP10,BFL10,SCP11,SSV12},
and ultracold atoms\,\cite{KFC09}) will soon allow to study and engineer the processes we have described 
here in laboratory.

\begin{acknowledgements}
  We would like to acknowledge valuable discussions with J.~Goold and
  R.~Dorner.  This project was support by Science Foundation Ireland
  under Project No.~10/IN.1/I2979.
\end{acknowledgements}



\begin{thebibliography}{99}
\bibitem{GB04} K.M.~Gaab and C.J.~Bardeen, J.~Chem.~Phys.~{\bf 121},
  7813 (2004).

\bibitem{OLO08} A. Olaya-Castro, C. Lee, F. Olsen, and N. Johnson, Phys. Rev. B, {\bf 78}, 085115 (2008).

\bibitem{NET10} P. Nalbach, J. Eckel, and M. Thorwart, New J. Phys., {\bf 12}, 065043 (2010).

\bibitem{NBT11} P. Nalbach, D. Braun, and M. Thorwart, Phys. Rev. E, {\bf 84}, 041926 (2011).

\bibitem{SMW11} T. Scholak, F. Melo, T. Wellens, F. Mintert, and A. Buchleitner, Phys. Rev. E, {\bf 83}, 021912 (2011).

\bibitem{RMK09} P.~Rebentrost, M.~Mohseni, I.~Kassal, S.~Lloyd, and
  A.~Aspuru-Guzik, New J.~Phys.~{\bf 11}, 033003 (2009).

\bibitem{PH08} M. Plenio and S. Huelga, New J. Phys., {\bf 10}, 113019 (2008).


\bibitem{GSN11a} P. K. Ghosh, A. Smirnov, and F. Nori, J. Chem. Phys. {\bf 134}, 244103 (2011).

\bibitem{GSN11b} P. K. Ghosh, A. Smirnov, and F. Nori, Phys. Rev. E {\bf 84}, 061138 (2011).


\bibitem{ECR07} G. S. Engel, T. R. Calhoun, E. L. Read, T. Ahn, T. Manal, Y. Cheng, R. E. Blankenship, and G. R. Fleming, Nature {\bf 446}, 782-786 (2007).  

\bibitem{PHF10} G. Panitchayangkoon, D. Hayes, K. A. Fransted, J. R. Caram, E. Harel, J. Wen, R. E. Blankenship,
and G. S. Engel, Proc. Natl. Acad. Sci., {\bf 107}, 12766 (2010).

\bibitem{CWW10} E. Collini, C. Y. Wong, K. E. Wilk, P. M. G. Curmi, P. Brumer, and
G. D. Scholes, Nature, {\bf 463}, 644 (2010).

\bibitem{RMA09} P. Rebentrost, M. Mohseni, and A. Aspuru-Guzik, J. Phys. Chem. B, {\bf 113}, 9942 (2009).


\bibitem{CCD09} F. Caruso, A. W. Chin, A. Datta, S. F. Huelga, and M. B. Plenio, J. Chem. Phys., {\bf 131}, 105106 (2009).

\bibitem{MRL08} M.~Mohseni, P.~Rebentrost, S.~Lloyd, A.~Aspuru-Guzik,
  J.~Chem.~Phys.~{\bf 129}, 174106 (2008).

\bibitem{Ria58}  G. V. Riazanov, Sov. Phys. JETP {\bf 6} 1107 (1958).

\bibitem{Fey86} R. Feynman, Found. Phys. {\bf 16}, 507 (1986).

\bibitem{Par88} K. R. Parthasarathy, Journal of Applied Probability, {\bf 25}, 151-166 (1988).

\bibitem{LP88} J. M. Lindsay and K. R. Parthasarathy, Sankhy$\bar{\rm a}$: The Indian Journal of Statistics, Series A, {\bf 50}, 151-170 (1988).

\bibitem{ADZ93} Y. Aharonov, L. Davidovich and N. Zagury, Phys. Rev. A {\bf 48}, 1687, (1993).

\bibitem{FG98} E. Farhi and S. Gutmann, Phys.Rev. A {\bf 58}, 915 (1998).

\bibitem{Kem03} J. Kempe, Contemp. Phys. {\bf 44}, 307 (2003).

\bibitem{Amb03} A. Ambainis, Int. Journal of Quantum Information, {\bf 1}, No.4, 507-518 (2003).

\bibitem{CL08} C. M. Chandrashekar and R. Laflamme,  Phys. Rev. A,  {\bf 78},  022314 (2008).

\bibitem{Cha11a} C. M. Chandrashekar, Phys. Rev. A, {\bf 83}, 022320 (2011).

\bibitem{KRB10} T. Kitagawa, M. S. Rudner, E. Berg, and E. Demler, Phys. Rev. A  {\bf 82},  033429 (2010).

\bibitem{CDE04} M. Christandl, N. Datta, A. Ekert, and A. J. Landahl, Phys. Rev. Lett., {\bf 92}, 187902 (2004).

\bibitem{KW11} P. Kurzy\'nski and A. W\'ojcik, Phys. Rev. A {\bf 83}, 062315 (2011).


\bibitem{GC10} S. K. Goyal, C. M. Chandrashekar, J. Phys. A: Math. Theor. {\bf 43},  235303 (2010).

 \bibitem{DGV12} R.~Dorner, J.~Goold, and V.~Vedral, Interface Focus
  rsfs20110109 (2012).


\bibitem{MB11} O.~ Muelken and A.~ Blumen  Physics Reports {\bf 502}, 37 - 87 (2011).

\bibitem{BCG04} E.~ Bach, S.~ Coppersmith, M.~ P.~Goldschen, R.~ Joynt, and J.~ Watrous, Journal of Computer and System Sciences,
Volume 69, Issue 4, Pages 562- 592, December 2004.

\bibitem{GAS11} M. G\"on\"ulol, E. Ayd�ner, Y. Shikano, and \"O. E.  M\"ustecaplio\"glu, New J. Phys. {\bf 13} 033037 (2011).


\bibitem{CSB07} C. M. Chandrashekar, R. Srikanth, and S. Banerjee, Phys. Rev. A, {\bf 76}, 022316 (2007).


\bibitem{BSC08} S. Banerjee, R. Srikanth, C. M. Chandrashekar, and P. Rungta, Phys. Rev. A, {\bf 78}, 052316 (2008).

\bibitem{LP10} C. Liu and N. Petulante, Phys. Rev. E, {\bf 81}, 031113 (2010).

\bibitem{L08} Luo, Phys. Rev. A {\bf 77} 022301 (2008).

\bibitem{SBC10} R. Srikanth, S. Banerjee, and C. M. Chandrashekar, Phys. Rev. A, {\bf 81}, 062123 (2010).


\bibitem{DLX03} J. Du, H. Li, X. Xu, M. Shi, J. Wu, X. Zhou, and R. Han, Phys. Rev. A {\bf 67}, 042316 (2003).

\bibitem{RLB05} C. A.  Ryan, M.  Laforest, J. C. Boileau, and R. Laflamme, Phys. Rev. A {\bf 72}, 062317 (2005).

\bibitem{LZZ10} D. Lu, J. Zhu, P. Zou, X. Peng, Y. Yu, S. Zhang, Q. Chen, and J. Du, Phys. Rev. A {\bf 81}, 022308 (2010).

\bibitem{SMS09} H. Schmitz, R. Matjeschk, Ch. Schneider, J. Glueckert, M. Enderlein, T. Huber, and T. Schaetz, Phys. Rev. Lett. {\bf 103}, 090504 (2009).

\bibitem{ZKG10} F. Zahringer, G. Kirchmair, R. Gerritsma, E. Solano, R. Blatt, and C. F. Roos, Phys. Rev. Lett. {\bf 104}, 100503 (2010).

\bibitem{PLP08} H. B. Perets, Y. Lahini, F. Pozzi, M. Sorel, R. Morandotti, and Y. Silberberg,  Phys. Rev. Lett. {\bf 100}, 170506 (2008).

\bibitem{SCP10} A. Schreiber, K. N. Cassemiro, V. Potocek, A. Gabris, P. Mosley, E. Andersson, I. Jex, and Ch. Silberhorn, Phys. Rev. Lett., {\bf 104}, 050502 (2010).

\bibitem{BFL10} M. A. Broome, A. Fedrizzi, B. P. Lanyon, I. Kassal, A. Aspuru-Guzik, and A. G. White. Phys. Rev. Lett. {\bf 104}, 153602 (2010).

\bibitem{PLM10} A. Peruzzo, M. Lobino, J. C. F. Matthews, N. Matsuda, A. Politi, K. Poulios, X. Zhou, Y. Lahini, N. Ismail, K. W\"orhoff, Y. Bromberg, Y. Silberberg, M. G. Thompson, and J. L. OBrien, Science {\bf 329}, 1500 (2010).

\bibitem{SCP11} A. Schreiber, K. N. Cassemiro, V. Potocek, A. Gabris, I. Jex, and Ch. Silberhorn, Phys. Rev. Lett. {\bf 106}, 180403 (2011).

\bibitem{SSV12} L. Sansoni, F. Sciarrino, G. Vallone, P. Mataloni, A. Crespi, R. Ramponi, and R. Osellame
Phys. Rev. Lett. {\bf 108}, 010502 (2012).

\bibitem{KFC09} K. Karski, L. Foster, J.-M. Choi, A. Steffen, W. Alt, D. Meschede, and A. Widera, Science  {\bf 325}, 174 (2009).



\end{thebibliography}
\end{document}